\documentclass{article} 
\usepackage{collas2022_conference,times}

\usepackage{amsmath,amsfonts,bm}

\def\eqref#1{equation~\ref{#1}}

\def\1{\bm{1}}

\DeclareMathAlphabet{\mathsfit}{\encodingdefault}{\sfdefault}{m}{sl}
\SetMathAlphabet{\mathsfit}{bold}{\encodingdefault}{\sfdefault}{bx}{n}

\usepackage{hyperref}
\hypersetup{
    colorlinks=true,
    linkcolor=red,
    filecolor=magenta,      
    urlcolor=blue,
    citecolor=purple,
    pdftitle={Overleaf Example},
    pdfpagemode=FullScreen,
    }
    
\usepackage{microtype} 
\usepackage{booktabs}
\usepackage[american]{babel}
\usepackage{amsmath, amssymb, natbib, graphicx, url}
\usepackage{mathtools} 
\usepackage{siunitx}
\usepackage{algorithm}
\usepackage[noend]{algpseudocode}
\usepackage{color, colortbl}
\usepackage{graphicx}
\usepackage{comment}
\usepackage{bibentry}
\usepackage{sectsty}
\usepackage{subcaption}
\usepackage{bm}

\title{PowerGraph: Using neural networks and principal components to determine multivariate statistical power trade-offs}

\author{Ajinkya K Mulay, Sean Lane, Erin Hennes \\
Department of Psychological Sciences \\
Purdue University \\
610 Purdue Mall \\
West Lafayette, Indiana, 47907 \\
\texttt{\{mulay, lane84, ehennes\}@purdue.edu} \\
}

\collasfinalcopy 

\begin{document}

\maketitle

\begin{abstract}
Statistical power estimation for studies with multiple model parameters is inherently a multivariate problem. Power for individual parameters of interest cannot be reliably estimated univariately since correlation and variance explained relative to one parameter will impact the power for another parameter, all usual univariate considerations being equal. Explicit solutions in such cases, especially for models with many parameters, are either impractical or impossible to solve, leaving researchers to the prevailing method of simulating power. However, the point estimates for a vector of model parameters are uncertain, and the impact of inaccuracy is unknown. In such cases, sensitivity analysis is recommended such that multiple combinations of possible observable parameter vectors are simulated to understand power trade-offs. A limitation to this approach is that it is computationally expensive to generate sufficient sensitivity combinations to accurately map the power trade-off function in increasingly high-dimensional spaces for the models that social scientists estimate. This paper explores the efficient estimation and graphing of statistical power for a study over varying model parameter combinations. We propose a simple and generalizable machine learning inspired solution to cut the computational cost to less than 10\% of the brute force method while providing F1 scores above 90\%. We further motivate the impact of transfer learning in learning power manifolds across varying distributions.
\end{abstract}

\section{Introduction}\label{sec:intro}

Statistical power is quantitatively equal to the probability of rejecting a null hypothesis. Thus, it plays a crucial role in finding an effect of hypothesized interest in any study. Historically, the simplest and most direct way to improve power is to increase the study's sample size, as mathematically, it has the most significant impact concerning what the researcher can control \cite{cohen1992things}. However, there are practical considerations to make when increasing the sample size. In the case of a rare disease study, the total population itself might be small, difficult to locate, and even more challenging to engage, as in early- and late-stage neuro-genetic syndromes \cite{button2013power, szucs2017empirical}. On the other hand, there could be access or funding issues with increasing the number of participants a researcher wishes to obtain.

\begin{figure}[htb!]
\begin{subfigure}{0.31\textwidth}
\includegraphics[width=\linewidth]{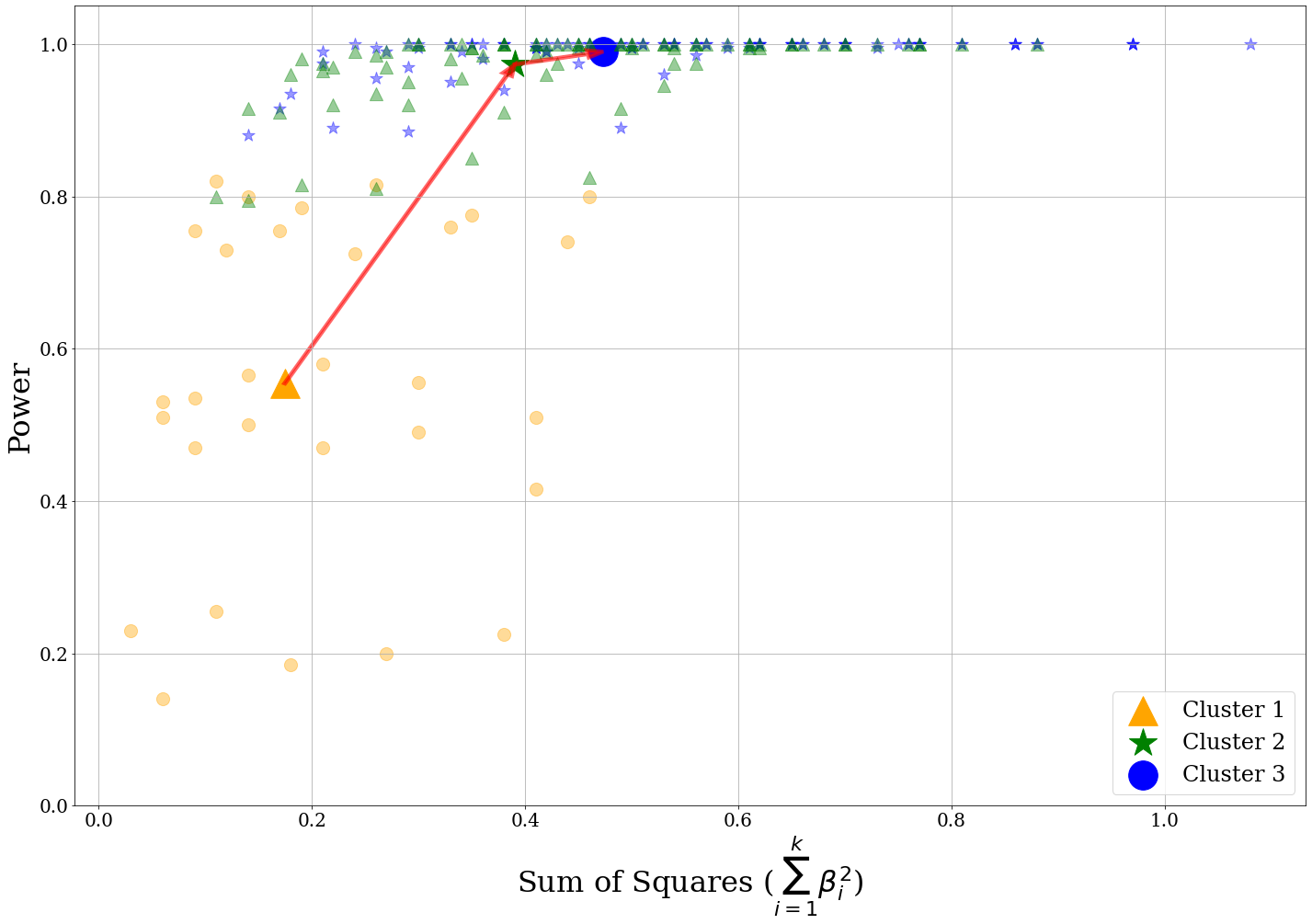}
\caption{} \label{fig:power-grad}
\end{subfigure}\hspace*{\fill}
\begin{subfigure}{0.31\textwidth}
\includegraphics[width=\linewidth]{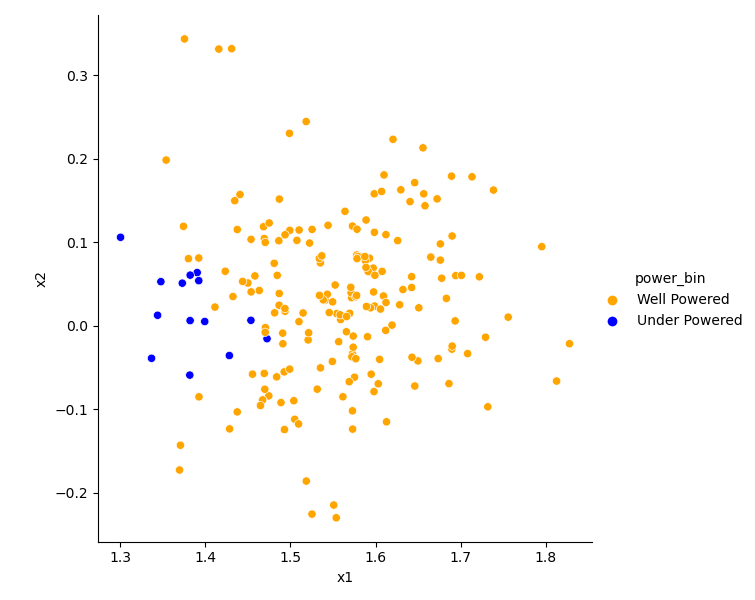}
\caption{} \label{fig:P11}
\end{subfigure}
\begin{subfigure}{0.31\textwidth}
\includegraphics[width=\linewidth]{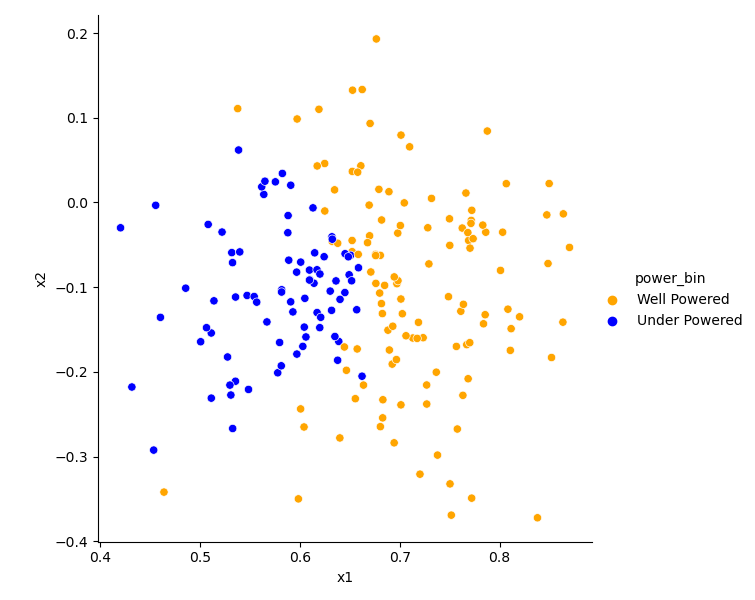}
\caption{} \label{fig:P21}
\end{subfigure}
\medskip
\begin{subfigure}{0.31\textwidth}
\includegraphics[width=\linewidth]{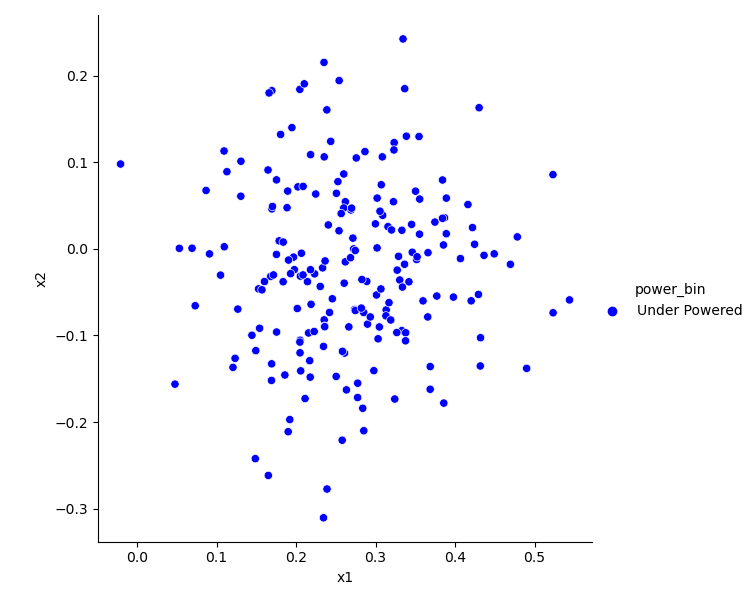}
\caption{} \label{fig:P31}
\end{subfigure}
\begin{subfigure}{0.31\textwidth}
\includegraphics[width=\linewidth]{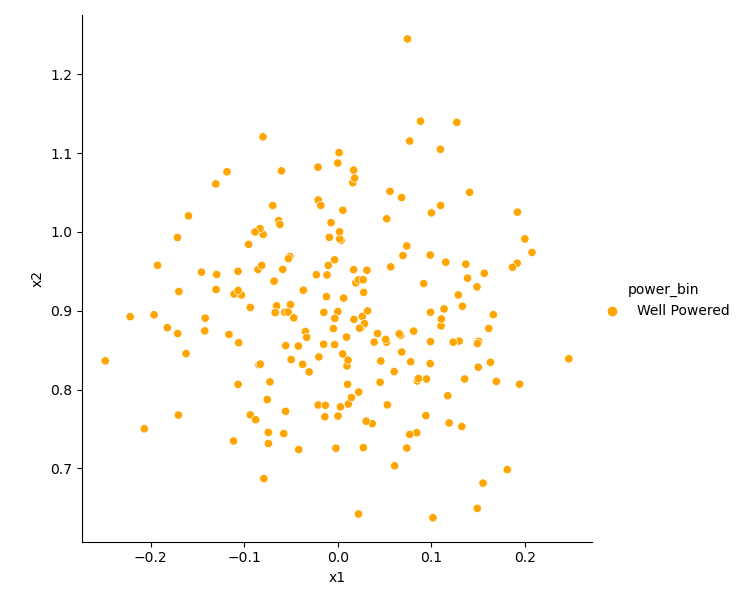}
\caption{} \label{fig:P41}
\end{subfigure}
\begin{subfigure}{0.31\textwidth}
\includegraphics[width=\linewidth]{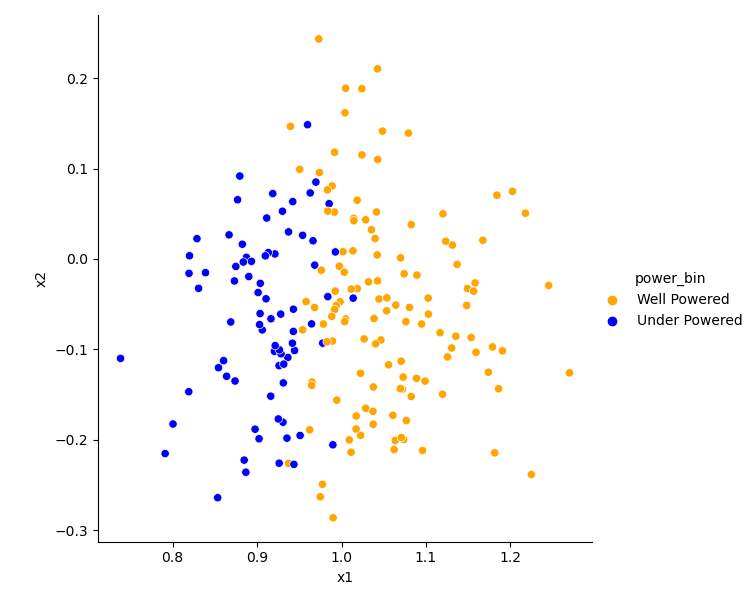}
\caption{} \label{fig:P51}
\end{subfigure}
\caption{\textbf{Impact of Model Parameters on Power: }\textit{(a)} Change in power over varying model coefficients ($\beta), N=105$. We observe significant power gradient over clusters derived by KMeans for a 3-predictor regression model with a partial f-test. \textit{(b, c, d, e, f)} For the same partial f-test and over popular regression datasets  \cite{kumarrajarshi_2015, Esoph, choi_2018, Toothgrowth, Parkinsons} ($N = 50$), we demonstrate that if we perturb the estimated model parameters ($\hat{\beta}$) by even small amounts, the under-powered studies can become well-powered (\emph{i.e.,} power $> 0.8$).}
\label{fig:model-param-power}
\end{figure}

In this article, we empirically show that tuning the model parameters (\emph{i.e.}, weights) can dramatically change the study's power. Thus, the change in model parameters can push the survey from a mid-powered study to well-powered research even for a constant sample size. For the rest of the article, we assume that well-powered research is one where the power is higher than \textit{0.8}. Fig.~\ref{fig:model-param-power} demonstrates that even for a fixed sample size, the power can considerab1y vary due to a change in model parameters. In Fig.~\ref{fig:power-grad} we compute power \textit{gradients} by considering the directional difference between clusters derived by KMeans, where the power and the squared L2-norm of the model parameters represent the point coordinates.

Thus, generating such a manifold, as shown in Fig.~\ref{fig:power-grad} can exceptionally aid researchers in identifying areas of high power. Enabling access to these plots is thus a simple way of reducing the \textit{ideal} sample size without sacrificing power. However, as described in Algorithm~\ref{algo:power}, computing power even once requires \textit{200-1000} simulations. Further, the manifold parameter space, including the model weights, sample size, and additional hyperparameter choices, is enormous. For instance, for a seven predictor model, with 11 choices per predictor and 11 choices for the $N$, we have a parameter space of $11^{7 + 1}$. Computation for this vast space might even take days.

Thus, in this work, we look at \textit{cheaper} alternatives to simulating the entire power manifold. We present three different contributions
\begin{itemize}
    \item We develop a neural network-based approach to predict power over a high-dimensional manifold for classification and regression purposes. Such an approach provides significant performance with even 10\% of the training data.
    \item We provide four baselines for alternatively predicting power and test all our approaches on three models-= regression, logistic and Repeated Measures ANOVA.
    \item To leverage the transfer of knowledge from one model's power manifold to another model's manifold, we first demonstrate the intuition behind such a transfer. Next, we show that such a transfer can enable the same or better results in every case.
We organize the rest of the article as follows. We first detail the relevant literature and highlight the background information required to build our algorithm. We provide pseudocodes for the core and auxiliary algorithms in this section. Next, we present our empirical results, highlight the significant wins, and report the limitations. Finally, we also present potential future work.
\end{itemize}

\section{Related Work}\label{sec:related-work}
Previous articles (\cite{bakker2012rules}, \cite{bakker2016researchers}, \cite{maxwell2004persistence}, \cite{cohen1992things}) have pointed out that studies are frequently underpowered and thus lead to statistically insignificant results. Underpowered studies are primarily a result of a lack of a formal power analysis. Furthermore, in \cite{card2020little} the authors demonstrated that numerous NLP models are typically underpowered. They mention that datasets with sentences less than 2000 lead to a power of only 75\% and provide best practices to perform power analyses. Next, the work in \cite{koehn2004statistical} provides ways to measure the statistical significance of the performance improvement due to a system change in the domain of machine translation. The authors argue that just a change in the system metric (ex., accuracy or F1-score) is not enough, and we need to demonstrate that the change in the metric is statistically significant. This work further broadens the need for thorough power analysis. Finally, \cite{koehn2004statistical} takes into consideration the effects of model complexity in extracting the maximum information from a given dataset. Under a given class of models, the authors provide an algorithm to identify the \textit{right-fit} model. The right-fit model avoids both underfitting and overfitting and has predictive power asymptotically close to the \textit{best} model in the given model class. Such a method can replace cross-validation and signifies the importance of improving predictive power.

\cite{bakker2012rules} suggested power simulations for studies using Questionable Research Practices (QRPs) \cite{john2012measuring} and lower sample sizes with more trials. QRPs could include running multiple tests with a much smaller sample size (underpowered), rerunning analyses after adding more subjects, or (even selectively) removing outliers. The results show that such practices can significantly inflate statistical power and provide misleading evidence about the effect size while hampering the study's reproducibility. \cite{bakker2012rules} suggests that to avoid such underpowered studies, we should use sample sizes derived after a formal power analysis.

\cite{baker2021power} solved the power issue by providing an online tool with power contours to demonstrate the effect of sample size (\textit{N}) and trials per participant on statistical power (\textit{k}). With the results provided in the article, it is clear that changes to \textit{N} or even \textit{k} can dramatically change the power region and convert an underpowered study to an appropriately powered one. \cite{rast2014longitudinal} also demonstrates the strong (inversely correlated) impact of sample size on both the effect size and study design. 

\cite{lane2018power} and  \cite{lane2019conducting} provide a clear guide to conducting formal power analysis based on simulation methods (rather than a formula-based approach). Even though the simulation approach is universal, it often requires many computational resources. The resource usage increases exponentially with a linear increase in the number of predictors or other factors impacting power. Thus, in this work, we highlight machine learning-based approaches that can seriously reduce resource usage with accuracies greater than 95\%.

\section{Proposed Work}

\subsection{Computing Power}

We present a power computation algorithm in Algorithm~\ref{algo:power} for t-tests in a linear regression model. We can extend this algorithm to the f-test by replacing the t-test in Line 9 with an f-test or a partial f-test. For extensions to other models, we only need to modify the data generation process on \textit{Line 5.} Computing power for multiple model weights requires us to call the \textit{COMPUTE-POWER} function each time. Thus, the cost of computing power boils down to the number of calls made to \textit{COMPUTE-POWER}. We wish to reduce the number of calls while still predicting power for the entire parameter space in our work.

\subsection{Feature Engineering with PCA}\label{sec:feat}

Complex datasets include several features, and it often becomes necessary to reduce data dimensionality to conserve resources or speed up training. Furthermore, we wish to remove redundancy in features. Removing highly correlated features can improve training speeds or data processing speeds, reduce bias, and improve the interpretability of our dataset. A common way to achieve these goals is to use Principal Component Analysis (PCA) \cite{wold1987principal}. For brevity, we add further details of PCA to the appendix. We also provide a simple PCA algorithm in Algorithm~\ref{algo:pca} in the appendix for completeness.

We already know that increasing the sample size $N$ can increase the power of a study. To visualize the impact of each parameter in the model, we take a look at a partial f-test. The data follows the distribution of the first three variables from Table~\ref{table:distribution-original}. We compute the power of the partial f-test to test for the significance of the first and third predictors. 

To visualize the importance of each parameter, we draw a correlation plot in Figure~\ref{fig:corr}. We can easily verify that as expected $\beta_{1}$ and $\beta_{3}$ have higher correlation with the power (as the hypothesis depends on them). $N$ has a high correlation as we would expect. Further, we define a new feature \textit{scaled weight} computed as $N\sigma$, wherein $\sigma = \sqrt{\sum_{i={1}}^{k} \beta_{i}^{2}}$ ($k$ denotes the number of predictors). We observe that the $N\sigma$ feature has the highest correlation amongst non-PCA features. 
After including the principal components the correlation substantially increases to $\geq 0.90$ for $\text{PC}_{1}$. We believe that transforming the data with PCA can further increase the variance while ignoring the redundant features in our dataset. We, provide exact details about our dataset in \ref{subsection:dataset}.

\subsection{Primary Approach: The Neural Network Approach}

We employ a simple neural network as described in Table~\ref{table:pnn-arch} trained on the \textit{new} dataset. With the flexibility of a neural network, we can train for both a classification and a regression task. We denote this approach by \textit{POWER-NETWORK} or \textit{POWER-Neural Network} (\textit{PNN})). \textit{PNN} also allows for an easy extension to a multi-class classifier. However, we do not explore this domain in this article.

\subsubsection{Learning Faster with Transfer Learning}

The primary goal of our article is to reduce the number of calls to the power function and thus reduce the simulation overhead for computing power. Thus, we now describe a commonly used technique \textit{fine-tuning} \cite{houlsby2019parameter} in Natural Language Processing (NLP) domains. Rather than cold-starting model training with random weights, we take the help of a previously trained model to initialize the model weights. Thus, most layers are already trained, and our new \textit{smaller} dataset can be used to tune the neural net for our new task. Note that when the number of input dimensions is different we remove the unused dimensions (from the pre-trained model) by specifying zero input columns for those features.

Formally, the transfer learning algorithm \textit{P-Transfer} only deviates from \textit{PNN} in the fact that line 6 of the algorithm~\ref{algo:power-network} has an additional step of initializing the model (before training) with the weights of a larger pre-trained model. Note that the original pre-trained model needs more features than the new model. Finally, the extra features in the larger model are turned off by passing zero vectors to these features. Thus, even a larger model can be used for our smaller models.

\subsubsection{Intuition behind Transfer Learning}

The intuition behind transfer learning may be sought from this article \cite{chinn2000simple}. We first note that each model directly interacts with the linear model. However, due to the difference in their composition, the effect size might differ. The authors in\cite{chinn2000simple} empirically demonstrate that given the standard deviation, we can find the correlation between, say, the effect size of the logistic regression model and the regression model. Thus, we would have the same standard deviation and thus correlated/comparable effect sizes for standardized predictors. Thus, we might be able to use transfer learning to boost a model with similar traits as long as the predictors are standardized and they both have the same hypothesis.

\subsection{Baselines: Power Cluster, Power Label Propagation and KNeighbors Classifier}

\textit{POWERCLUSTER} also referred to as \textit{P-CLUSTER}) simply runs K-Means on $PC_{1}$ and $N\sigma$ for identifying two clusters with two vectors from our dataset. We provide its pseudocode in Algorithm~\ref{algo:power-cluster}. K-Means identifies clusters of points that minimize the intra-cluster distances. Since both of our feature vectors are independent of power, we can cluster our data points without computing the \textit{true} power.

\begin{figure}[htb!]
\begin{subfigure}{0.48\textwidth}
\includegraphics[width=\linewidth]{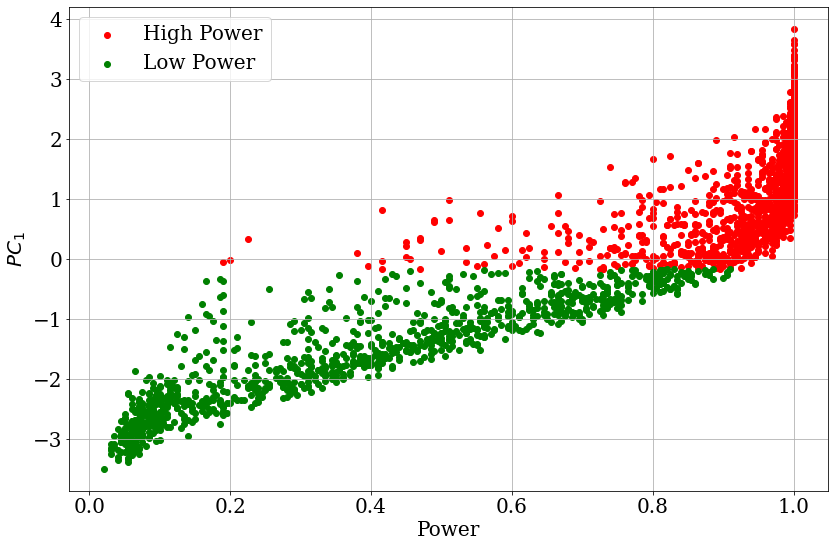}
\caption{} \label{fig:power-grad2}
\end{subfigure}\hspace*{\fill}
\begin{subfigure}{0.48\textwidth}
\includegraphics[width=\linewidth]{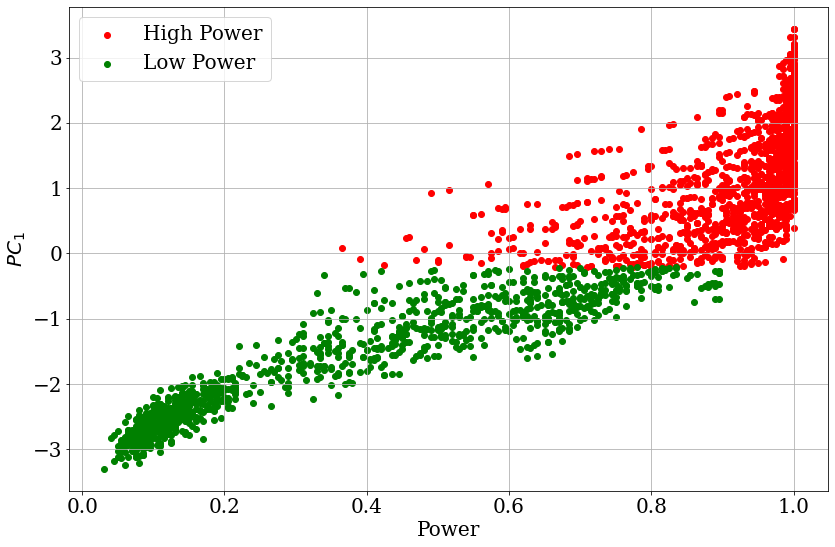}
\caption{} \label{fig:P12}
\end{subfigure}
\caption{\textbf{PNN-CLUSTER Performance: }A simple unsupervised learning algorithm can identify high and low power domains. \textit{(left)} Results for partial f-test performed on a 3-predictor \textit{REG} model. \textit{(right)} Results for partial f-test performed on a 5-predictor \textit{REG} model.}
\label{fig:power-cluster-plot}
\end{figure}

The intuition behind deploying \textit{P-CLUSTER} is that it can quickly identify appropriately varying power domains, as seen from Fig.~\ref{fig:power-cluster-plot}. If we compare the clustering to the standard definition of a high-powered study (\emph{i.e., } power is more significant than \textit{0.8}), we can segregate power with high performance. We demonstrate the results in Figures~\ref{fig:r8}, \ref{fig:r6}. We can see that \textit{P-CLUSTER} performs well arbitrarily. Thus, we can conclude that not including label information will lead to poor performance for models.

 We provide two other strong baselines for power prediction- the label propagation algorithm from \cite{zhu2002learning}, and the KNeighbors Classifier \cite{nelli2018machine}. We use the Label Propagation version implemented in Scikit-Learn \cite{scikit-learn}. Label propagation works by first creating a fully-connected graph between all data points wherein each edge weight is directly dependent on the euclidean distance between the two points. These edge weights dictate the probability of propagating a label onto another label. We direct the reader to section 2.2 of \cite{zhu2002learning} for further details on the Label Propagation algorithm. We also refer to it as \textit{PL-PROP}.

KNeighbors classifier \cite{nelli2018machine} also provides good performance, and we thus include it for completeness. Overall, Label Propagation performs consistently better than the KNeighbors classifier.

\begin{algorithm}[htb!]
\caption{Computing Power of a t-test for a Linear Regression Model}
\label{algo:power}
\begin{algorithmic}[1]
\State \textbf{Input:} Distributions of columns in the dataset-$\mathcal{D}=\{D_{X_{1}}, D_{X_{2}}, ..., D_{X_{p}}\}$, Sample Size-$N$, Model Weight-$\beta \in \mathbb{R}^{p}$, Number of predictors-$p$, sensitivity-$\alpha$ (\textit{default:} 0.05), number of simulations-\textit{sims} (\textit{default: 1000}), Error Distribution-$\mathcal{E}$ 
\State \textbf{Output:} Power of t-test
\Procedure{COMPUTE-POWER}{$X, k$}
\State \textit{significance} $\leftarrow$ 0
\For{1:\textit{sims}}
    \State $X \leftarrow$ generate $N$ data samples from 
    \State distribution($\mathcal{D}$)
    \State $e \leftarrow$ generate error from distribution($\mathcal{E}$)
    \State $y \leftarrow (X \times \beta) + e$
    \State $\mathcal{M} \leftarrow$ Fit $(X, y)$ to a linear regression model
    \If{p-value of t-test for model $\mathcal{M} \leq \alpha$}
        \State \textit{significance} += 1
    \EndIf 
\EndFor 
\State power $\leftarrow \frac{\textit{significance}}{\textit{sims}}$
\State return power
\EndProcedure
\end{algorithmic}
\end{algorithm}

\begin{algorithm}[htb!]
\caption{Unsupervised clustering of the power surface with PCA features}
\label{algo:power-cluster}
\begin{algorithmic}[1]
\State \textbf{Input:} Length of training set-$L$, model parameter space-$\mathcal{S}$ of length $L$ (each parameter consists of the weight vector $\beta$ and sample size $N$), power sensitivity-$\alpha$ (\textit{default 0.05}), number of simulations-\textit{sims} (\textit{default 200}), PCA variance (in \%) to be retained-\textit{var}, Number of clusters-$k$
\State \textbf{Output:} Power Surface Graph
\Procedure{POWERCLUSTER}{$\mathcal{S}, \alpha$, \textit{sims}}
\State $\mathcal{S}_{\text{PCA}}$ $\leftarrow$ PCA-FIT-TRANSFORM(S, variance=\textit{var})
\State $C_{k}$ $\leftarrow$ KMeans($\mathcal{S}_{\text{PCA}}$, num\_clusters $= k$)
\State return clusters $C_{k}$ derived from KMeans
\EndProcedure
\end{algorithmic}
\end{algorithm}

\begin{algorithm}[htb!]
\caption{Training Data Collection}
\label{algo:collect-training-data}
\begin{algorithmic}[1]
\State \textbf{Input:} Length of training set-$L$, model parameter space-$\mathcal{S}$ of length $L$ (each parameter consists of the weight vector $\beta$ and sample size $N$), power sensitivity-$\alpha$ (\textit{default 0.05}), number of simulations-\textit{sims} (\textit{default 200})
\State \textbf{Output:} Training Set $\mathcal{S}$ with parameters and corresponding powers
\Procedure{GENERATE-DATA}{$\mathcal{S}, \alpha$, \textit{sims}}
\For{parameter in $\mathcal{S}$}
    \State ($\beta, N) \leftarrow$ parameter
    \State power $\leftarrow$ COMPUTE-POWER($\beta,N,\alpha,$\textit{sims})
    \State $\mathcal{S}$[powers] $\leftarrow$ power
\EndFor
\EndProcedure
\end{algorithmic}
\end{algorithm}

\begin{algorithm}[htb!]
\caption{Classifying/Predicting power surface with PCA features}
\label{algo:power-network}
\begin{algorithmic}[1]
\State \textbf{Input:} Length of training set-$L$, model parameter space-$\mathcal{S}$ of length $L$ (each parameter consists of the weight vector $\beta$ and sample size $N$), power sensitivity-$\alpha$ (\textit{default 0.05}), number of simulations-\textit{sims} (\textit{default 200}), PCA variance (in \%) to be retained-\textit{var}
\State \textbf{Output:} Power Surface Graph
\Procedure{POWERNETWORK}{$\mathcal{S}, \alpha$, \textit{sims}}
\State $\mathcal{S}^{*} \leftarrow$ GENERATE-DATA($\mathcal{S}, \alpha$, \textit{sims}) 
\State $\mathcal{S}_{\text{PCA}}$ $\leftarrow$ PCA-FIT-TRANSFORM($\mathcal{S}$, variance=\textit{var})
\State Train Neural Network Model $\mathcal{M}$ on ($\mathcal{S}^{*} \cup \mathcal{S}_{\text{PCA}}$)
\State return trained model $\mathcal{M}$
\EndProcedure
\end{algorithmic}
\end{algorithm}

\section{Experiments}

\subsection{Code}\label{sec:code}

We test the efficacy of the baseline algorithms and \ref{algo:power-network} with three kinds of statistical models. For simulation, we use Google Colaboratory with the standard run-time.\footnote{Our code is available at the anonymized link \url{https://anonymous.4open.science/r/powergraph-4BFC}.} Other experiment details are deferred to the appendix in section~\ref{section:experiments}.

\begin{algorithm}[htb!]
\caption{Sampling in High-Dimensions}
\label{algo:p-sampler}
\begin{algorithmic}[1]
\State \textbf{Input:} no. of training points-$N_{s}$, domain of the $k$ predictors and the sample size-$\mathcal{D} \in \mathbb{R}^{k}$, no. of local training points-$n_{s}$, local sigma for Gaussian sampling-$\sigma_{l}$
\State \textbf{Output:} $\mathcal{S}$ model parameter space
\Procedure{P-SAMPLER}{$N_{s}$}
\State $n \leftarrow 0$
\While{$n < N_{s}$}
    \State Uniformly randomly pick a centroid $\beta$ from $\mathcal{D}$
    \State $\mathcal{S}_{l} \leftarrow $ Sample $n_{s}$ datapoints from a Gaussian distribution $\mathcal{N}(\mu=\beta, \sigma=\sigma_{l})$
    \State $\mathcal{S} \leftarrow \mathcal{S} \cup \mathcal{S}_{l}$
    \State $n \leftarrow n + 1 + n_{s}$
\EndWhile
return parameter space $\mathcal{S}$
\EndProcedure
\end{algorithmic}
\end{algorithm}

\subsection{Dataset}\label{subsection:dataset}

For sampling parameters we use \textit{P-SAMPLER} (algorithm~\ref{algo:p-sampler}) to generate the parameter sample space and in conjunction with algorithm~\ref{algo:collect-training-data} to compute power. Both of these algorithms together make up the power computation black box. After collecting the dataset, we pre-process it using PCA, as explained below. \textit{P-SAMPLER} is inspired by the fungible weights described in \cite{waller2008fungible}. 
Next, we provide exact details about the data we use for the methods introduced in this work. For both supervised and unsupervised tasks we use the  same dataset albeit the unsupervised one does not contain the labels or true power information. The dataset details are included in Table~\ref{table:dataset}. We assume that our original linear model is denoted as 
\begin{align*}
    y = x_{1} \beta_{1} + x_{2} \beta_{2} + ... + x_{k} \beta_{k} + e
\end{align*}
wherein, $e \sim \mathcal{N}(0, 1)$ is the error term, $k$ is the number of predictors, $y$ is the model predictions, $x = \{x_{1}, ..., x_{k}\}$ and $\beta = \{\beta_{1}, ..., \beta_{k}\}$ is our linear model parameter. Our modified dataset includes $X = \{\beta, N, N\sigma\}$. After including the PCA induced features our final dataset is $X_{PCA} = \{\beta, N, N\sigma, \text{PC}\}$.

\subsubsection{Models}

Three models are considered for understanding how the power manifold is impacted by model complexity. $x$ below signifies the number of predictors in the model.
\begin{itemize}
    \item \textit{Linear Model (REG-x):} We run an f-test to understand whether a simpler model works better than the full model. Feature distribution listed in Table~\ref{table:distribution-original}.
    \item \textit{Logistic Regression (LOGIT-x):} Here, we run a Wald's test to understand how much a reduced model impacts the predictive power. This is the only non-linear model that we consider. Feature distribution listed in Table~\ref{table:distribution-alternative}.
    \item \textit{Repeated Measures ANOVA (RMANOVA-x):} We run a within-subjects RMANOVA to measure the impact of different factors (two) on the subjects. Feature distribution listed in Table~\ref{table:distribution-original}.
\end{itemize}

Throughout the paper, we use the same hypothesis $\mathcal{H}_{\mathcal{O}}$ where we test whether the model parameters $\beta_{1}, \beta_{3}$ are zero (in the form of a partial F-test or a Wald's test). However, for testing the robustness of our transfer learning algorithm \textit{P-TRANSFER}, we use the alternative hypothesis $\mathcal{H}^{'}_{\mathcal{O}}$ where we test whether the model parameters $\beta_{1}, \beta_{7}, \beta_{8}$ are zero or not for both an F-test and a Wald's test.

\begin{table}[htb!]
\begin{center}
\begin{tabular}{|c|c|} 
 \hline
 \rowcolor{lightgray}
 Feature & Details \\ 
 \hline \hline
 $\beta$ & As provided by the expected dataset's distribution \\
 \rowcolor{lightgray}
 $N, N\sigma$ & Sample Size, Scaled Weight \\ 
 $\text{PC} = \{\text{PC}_{1}, ..., \text{PC}_{r}\}$ & $r$ Principal Components as extracted from $X$ with 99\% variance\\ 
 \hline
\end{tabular}
\end{center}
\caption{Dataset: Features included for model training}
\label{table:dataset}
\end{table}

For the \textit{3}-predictor network we have \textit{9} total features, \textit{13} features for the \textit{5}-predictor network, \textit{23} for the \textit{10}-predictor network and \textit{43} for the \textit{20}-predictor network.


\subsection{Metrics}

In our proposed algorithm, we solve two kinds of problems- binary power classification and a regression problem with actual power values. 
\vspace{-2mm}
\subsubsection{Classification}
\vspace{-1mm}
For binary classification, we normally simply use the he accuracy.
However, since we might have unbalanced datasets, we instead report the F1 score over the test set.
While using multiple binary classifiers we could still use the F1-score albeit in an \textit{one-vs-rest} fashion. For multi-class classification, we are first required to run the more restrictive binary classifier, compute its performance, and then for the second binary classifier we only compute its performance over the remaining data points. Thus, suppose we have two binary classifiers- $C_{1}$ (\textit{power} $> 0.8$ and \textit{power} $\leq 0.8$) and $C_{2}$ (\textit{power} $> 0.6$ and \textit{power} $\leq 0.6$). We first run $C_{1}$ to find all data points with \textit{power} $> 0.8$ and then run $C_{2}$ to find data points with $0.8 \geq $\textit{power} $> 0.6$ and \textit{power} $\leq 0.6$. Note that currently in the results we only provide separate performance metrics for $C_1$ and $C_2$.
\vspace{-2mm}
\subsubsection{Regression}
\vspace{-1mm}
We use a Mean Squared Error loss function for the regression problem between the ground truth power and the simulated power values. However, power is the probability of successfully rejecting a null hypothesis, and thus we can consider the ground truth power $P^{g}$ and the simulated power values $P^{s}$ as probability distributions. Note that the KL Divergence between $A$ and $B$ is shown in equation~\ref{eqn:KL-divergence}.
\begin{equation}\label{eqn:KL-divergence}
    \mathcal{L}_{KL}(A || B) = \sum_{x} A(x) \text{log} \Big( \frac{A(x)}{B(x)} \Big)
\end{equation}
Here, $x$ denotes the feature space associated with the probability distribution (\emph{i.e., } actual and simulated power), and $A, B$ represents the ground truth and the simulated powers themselves.
However, KL divergence is non-symmetric and not a distance measure. Thus, we use the Jensen-Shannon (JS) divergence which is an extension of the equation.~\ref{eqn:KL-divergence}. The JS divergence between two probability distributions $A$ and $B$ is given by the equation~\ref{eqn:JS-divergence}.
\begin{equation}\label{eqn:JS-divergence}
    \mathcal{L}_{JS}(A || B) = \frac{1}{2} \{\mathcal{L}_{KL}(A || B) + \mathcal{L}_{KL}(B || A)\}
\end{equation}

\subsection{A note about current baselines for predicting power}

From literature, previous approaches for computing a power manifold used a brute-force approach by essentially running the \textit{black box} for computing power. Thus, given unlimited time, they could achieve 100\% accuracy by just literally computing the power manifold. Specific greedy approaches attempted to find optimal power but not the entire manifold. Since our work aims to minimize the effort required to compute the power manifold, we do not compare our work directly to these brute-force approaches. Instead, we elect to create our baselines using traditional predictive machine learning.

\section{Results}\label{sec:results}

\subsection{P-Transfer}

We evaluate P-Transfer by first pre-training a \textit{REG-20} model and then testing it over three different scenarios-(i) Same feature distribution, Same Hypothesis, (ii) Different feature distribution, Same Hypothesis, and (iii) Different feature distribution, Different Hypothesis. We denote these cases by $TF_{1}, TF_{2}, TF_{3}$ respectively. Note that we always use a model smaller or equal in size to the original pre-trained model for transfer learning. 

Our preliminary observations indicate that transfer learning \textit{never} hurts model performance. However, for the cases where both the models have the same hypothesis (\emph{i.e., } cases (i) and (ii)), we see significant gains in the model performance. Table~\ref{table:p-transfer} summarizes these results.

\begin{table}[htb!]
\begin{center}
\begin{tabular}{|p{1cm}|p{3cm}|p{3cm}|p{3cm}|p{3cm}|} 
 \hline
 \rowcolor{lightgray}
 Type & Model & P-Transfer Performance & Original Performance (Same Amt. of Train Data) & Original Performance (8x Amt. of Train Data) \\ 
 \hline \hline
 $TF_{1}$ & REG3 & $\bm{0.9902}^{\dagger} \bm{\pm 0.0045}$ & $0.9827$ & $0.9900$ \\
 $TF_{1}$ & REG5 & $\bm{0.9804}^{\dagger} \bm{\pm 0.0064}$ & $0.9700$ & $0.9800$ \\
 $TF_{1}$ & REG10 & $\bm{0.9630 \pm 0.0087}$ & $0.9600$ & $0.9800$ \\
 \rowcolor{lightgray}
 $TF_{1}$ & RMANOVA3 & $\bm{0.9458 \pm 0.0104}$ & $0.945$ & $0.9600$ \\
 \rowcolor{lightgray}
 $TF_{1}$ & RMANOVA5 & $\bm{0.9650 \pm 0.0084}$ & $0.9600$ & $0.9700$ \\
 \rowcolor{lightgray}
 $TF_{1}$ & RMANOVA10 & $\bm{0.9465 \pm 0.0103}$ & $0.9200$ & $0.9800$ \\
 \rowcolor{lightgray}
 $TF_{1}$ & RMANOVA20 & $0.8513 \pm 0.0164$  & $\bm{0.8600}$ & $0.9300$ \\
 $TF_{2}$ & LOGIT3 & $\bm{0.9717}^{\dagger} \bm{\pm 0.0076}$ & $0.9300$ & $0.9700$ \\
 $TF_{2}$ & LOGIT5 & $\bm{0.9524 \pm 0.0098}$ & $0.9200$ & $0.9600$ \\
 $TF_{2}$ & LOGIT3 & $\bm{0.9361 \pm 0.0112}$ & $0.8900$ & $0.9700$ \\
 $TF_{2}$ & LOGIT3 & $\bm{0.9099 \pm 0.0132}$ & $0.8600$ & $0.9300$ \\
 \rowcolor{lightgray}
 $TF_{2}$ & REG10 & ${0.9658 \pm 0.0178}$ & $\bm{0.9659}$ & $0.9944$ \\
 \rowcolor{lightgray}
 $TF_{2}$ & RMANOVA10 & $\bm{0.9260 \pm 0.0256}$ & $0.8678$ & $1.0000$ \\
 $TF_{3}$ & RMANOVA20 & $\bm{0.815 \pm 0.0380}$ & $\bm{0.815}$ & $0.815$\\ \hline
\end{tabular}
\end{center}
\caption{\textit{P-Transfer Test Performance}: Models with the same hypothesis may provide considerable performance boost. Any other scenario does not seem to hurt transfer learning performance.$\dagger$-indicates that P-Transfer beat even the 8x training data performance. The confidence ranges indicate a 95\% confidence interval.}
\label{table:p-transfer}
\end{table}

Thus, for comparison to \textit{PNN}, we use our baselines -\textit{PL-Rand, P-CLUSTER, PL-PROP} and the \textit{PK-neighbors} approach. Note that to accommodate the various trends in our model, we only report the best performing baseline (over all four baselines) in the plots. For brevity and due to many baseline performance combinations, we skip the point-wise results of the baselines. However, the code provided hosts with all the information required to recreate these baselines.

We capture three primary trends in our experiments, the performance over the change in model complexity, the number of predictors, and the amount of training data used for predicting the power manifold.

\begin{figure}[htb!]
\begin{subfigure}{0.48\textwidth}
\includegraphics[width=\linewidth]{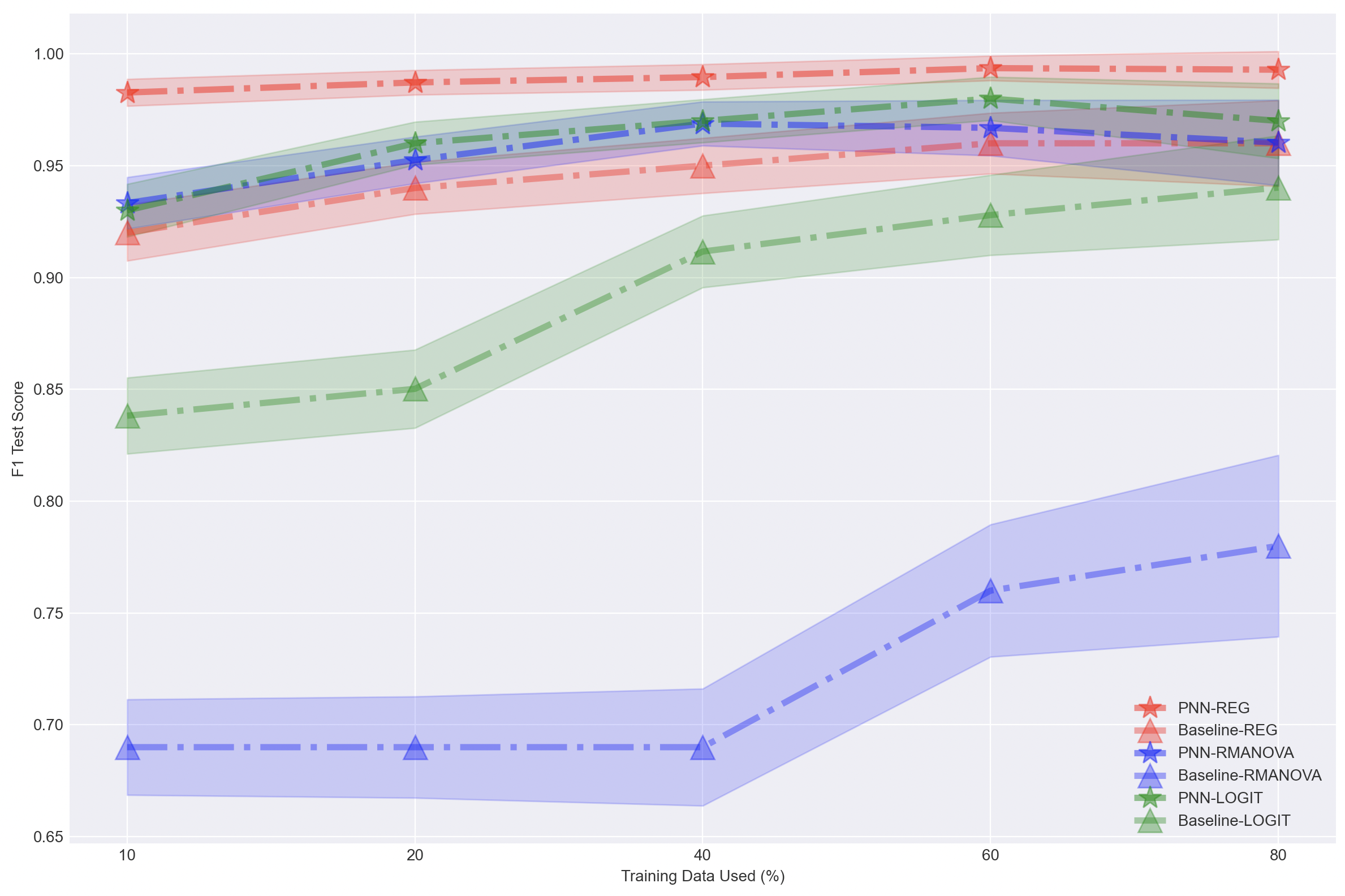}
\caption{} \label{fig:r81}
\end{subfigure}\hspace*{\fill}
\begin{subfigure}{0.48\textwidth}
\includegraphics[width=\linewidth]{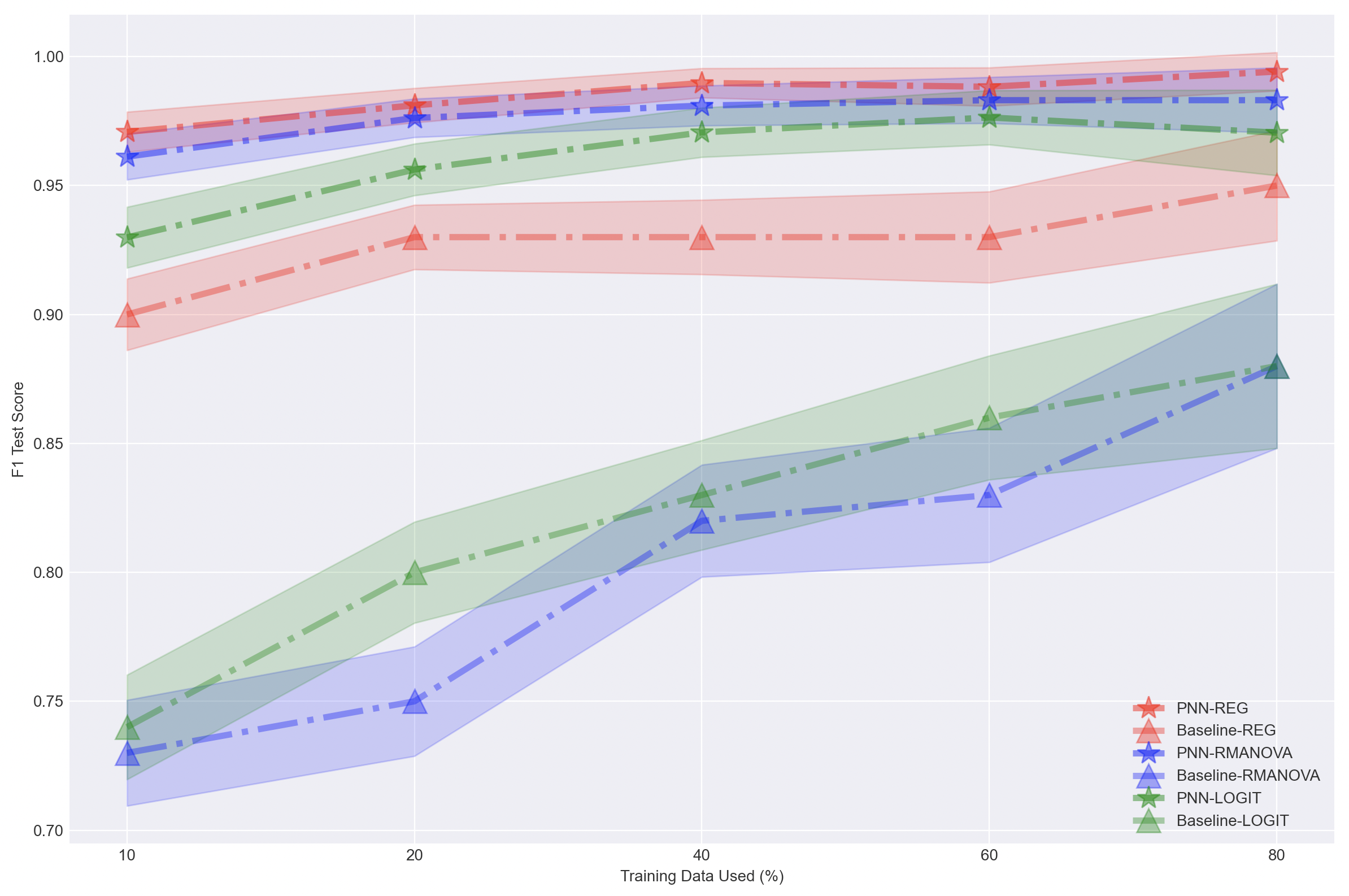}
\caption{} \label{fig:r61}
\end{subfigure}
\caption{\textbf{PNN Classification Performance:} Comparison of model performance over change in proportion of training data used (increasing from left to right). \textbf{(left)}\textit{} 3-predictor performance, and \textbf{\textit{(right)}} 5-predictor network performance.}
\label{fig:mainR}
\end{figure}

\begin{figure}[htb!]
\begin{subfigure}{0.48\textwidth}
\includegraphics[width=\linewidth]{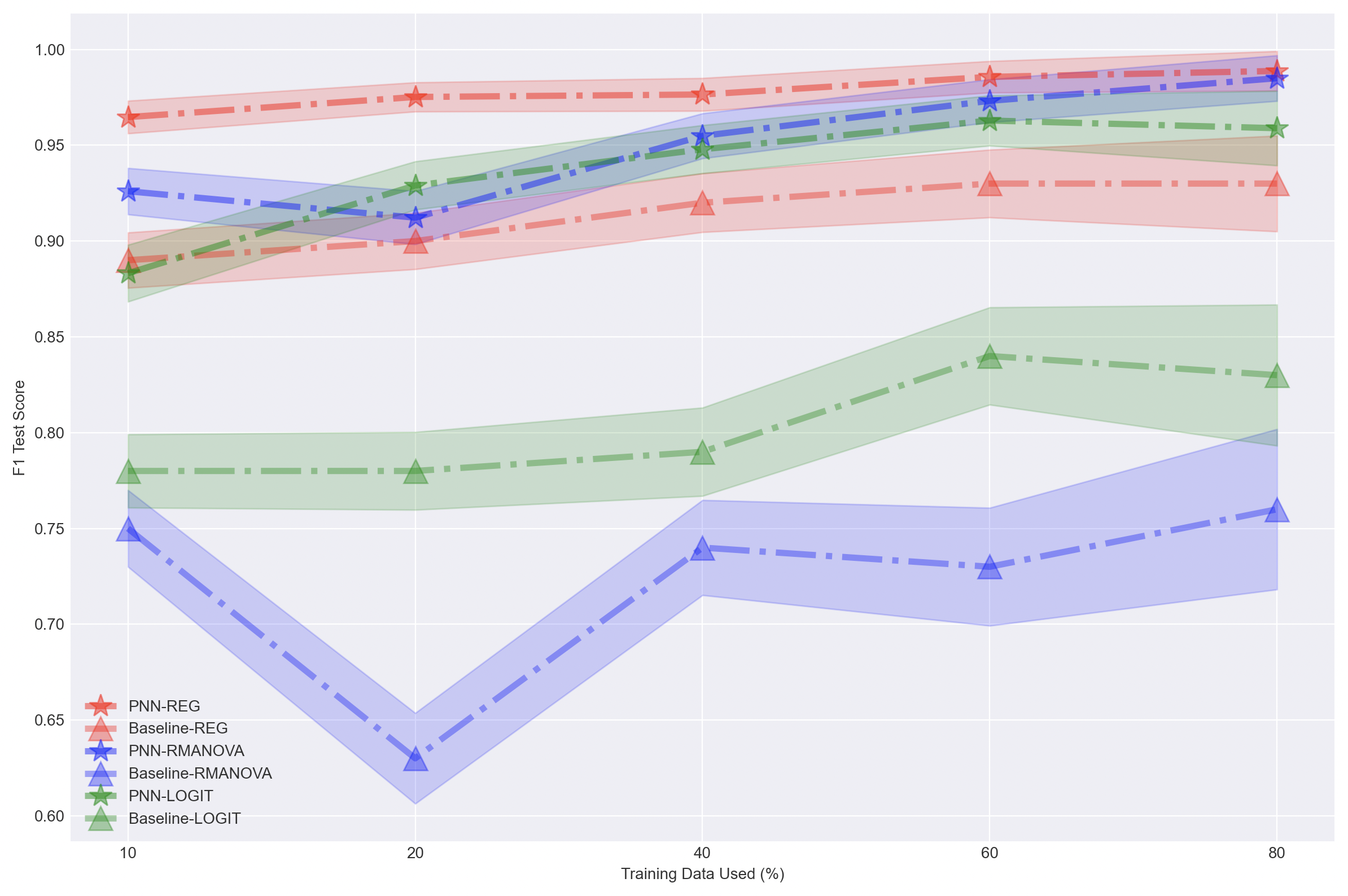}
\caption{} \label{fig:r8}
\end{subfigure}\hspace*{\fill}
\begin{subfigure}{0.48\textwidth}
\includegraphics[width=\linewidth]{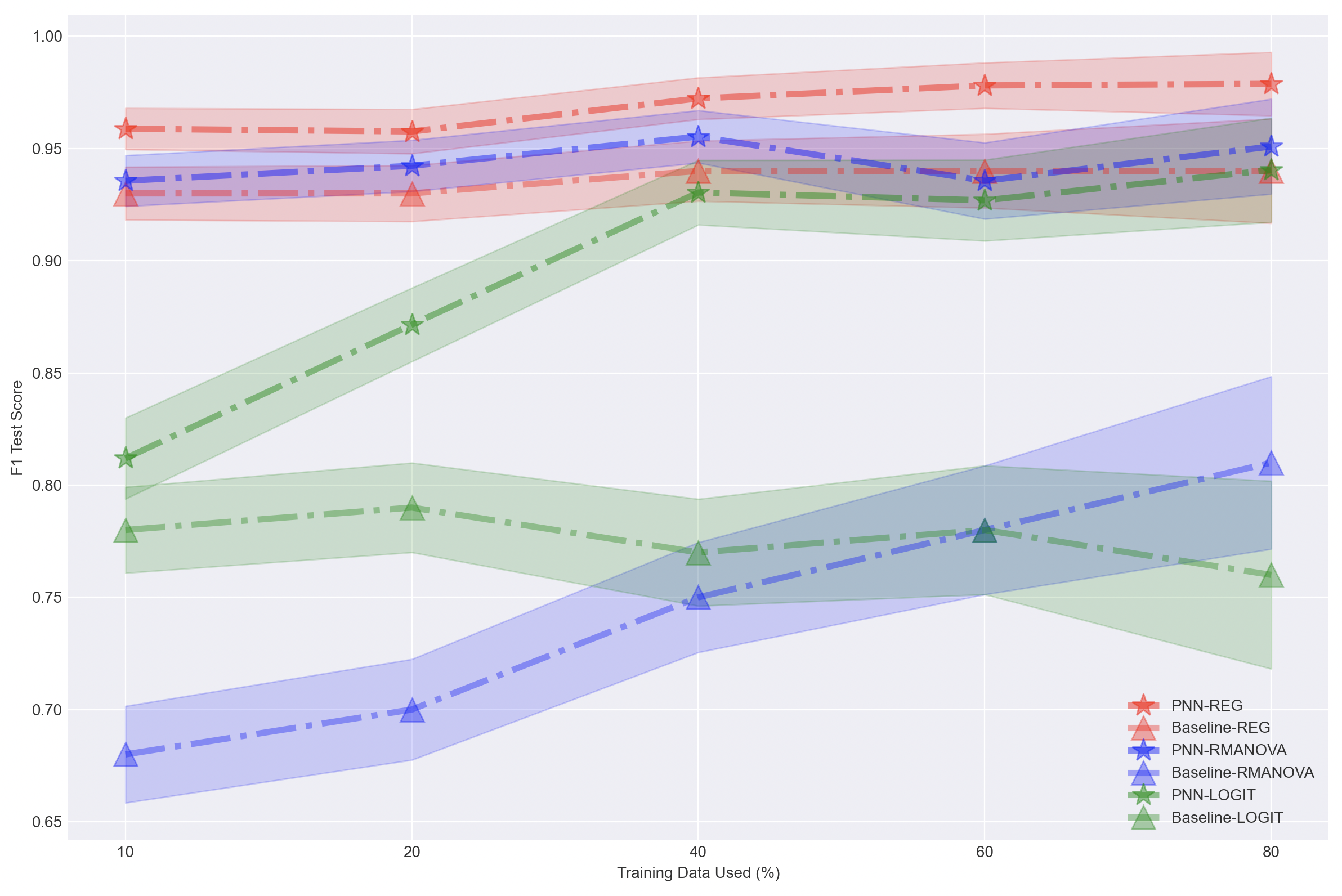}
\caption{} \label{fig:r6}
\end{subfigure}
\caption{\textbf{PNN Classification Performance:} Comparison of model performance over change in proportion of training data used (increasing from left to right). \textbf{(left)}\textit{} 10-predictor performance, and \textbf{\textit{(right)}} 20-predictor network performance.}
\label{fig:mainA}
\end{figure}

The trend of increased performance with increased training data usage is reported via Figures~\ref{fig:mainA}, \ref{fig:mainR}. Interestingly, even at lower training data proportions, we receive good performance from the classifiers. From the exact figures, we can see that predictions struggle over the RMANOVA model while they are much easier for the REG model and moderate for the LOGIT model. A similar trend can be observed for regression tasks as seen in the Figs.~\ref{fig:REGR}, \ref{fig:RMANOVAR}, \ref{fig:LOGITR} in the appendix. Furthermore, with a change in the number of predictors as seen in Figs.~\ref{fig:classREG}, \ref{fig:classRMANOVA} and \ref{fig:classLOGIT}, the model performance deteriorates for change in the number of predictors for REG and LOGIT. However, for the RMANOVA model, we see an arbitrary performance plot  indicating our sub-optimal model choice might be the reason. Here sub-optimal indicates the model derived without using the validation set. It seems that for both regression and classification tasks, PNN struggles the most for the RMANOVA model family. However, given that we see this is only the models with low access to training data, we can say that these classifiers are biased.

Finally, we can observe the number of calls v/s the performance in Fig.~\ref{fig:comp-costs}. Clearly, for multiple models using fewer data points can still provide high utility.

Even though we do not fully report the baselines, we would like to highlight a few notable observations. As expected, \textit{P-RAND} performs the worst, followed by \textit{P-CLUSTER}. Depending on the simulated data's distribution \textit{P-CLUSTER} surprisingly beats both \textit{PK-neighbors} and \textit{PL-PROP} when fewer training data points are used. Amongst, \textit{PL-PROP} and \textit{PK-neighbors}, the \textit{PL-PROP} approach is the more robust one and it is the closest one to \textit{PNN}.

\section{Limitations and Future Work}

This work primarily investigates the utility of training a neural network for predicting statistical power over a manifold. Furthermore, we also provide baselines- Random (P-Rand), K Neighbors Classifier (PK-Neighbors), Label Propagation (PL-Prop), and an unsupervised clustering method (P-CLUSTER).

Note that each of the baselines fails to generalize for complex models.
\begin{itemize}
    \item \textit{P-RAND}: The random mechanism works as advertised and provides the least performance throughout. 
    \item \textit{P-CLUSTER}: This approach can be compelling for visualizing the power manifold in 2 dimensions space effectively. However, we note that its performance is inconsistent, and the linear classifier might not always be able to predict exactly between high and low power regions. Further, the lack of label information makes this approach useful for pre-processing but not for predicting power classes. Alternatively, we might be able to identify and eliminate \textit{} low powered regions using this approach, and we again propose this as an open problem. 
    \item \textit{PK-neighbors}: This approach is a supervised clustering alternative to \textit{P-CLUSTER}. However, it does not necessarily perform consistently. We, however, note that we have not fine-tuned this approach for the number of neighbors. We leave this investigation as part of our future work.
    \item \textit{PL-PROP}: The most promising baseline is \textit{PL-PROP} with its \textit{rbf} kernel. \textit{PL-PROP} is better than the rest but fails to generalize for complex models or more predictors. Furthermore, it requires much more data compared to the neural network. We, however, have not used any guidance for tuning this algorithm's parameters, and we leave this aspect for future work.
    \item \textit{Model Limitations:} \textit{PNN} provides excellent empirical performance and generalizes well over increasing model complexity while providing competitive performance even with only 10\% of the data. We note that \textit{PNN} is not yet trained using methods like \textit{early stopping} or an \textit{adaptive learning rate}. We would pursue these optimizations in subsequent work. Particularly we do not use a validation dataset to improve the model generalization, and we set \textit{500} epochs as an arbitrary stopping point. We have seen at least one instance of such an approach leading to a sub-optimal model choice, thus leaving a gap in the empirical performance.
    \item \textit{Dataset Limitations:} Our most complex model currently is the RMANOVA with two within-subjects factors. To further understand the generalization error, we hope to extend our approach to higher complexity models, including other non-linear models. Further, we would like to explore multi-class classifications to provide the user with more options in terms of choosing the power manifold.
    \item \textit{Theoretical Limitation:} Finally, the current implementations do not provide any formal theoretical guarantees for the convergence of any of these methods. We do provide intuition about why our transfer learning methods work. However, we would like to explore how different distributions with the same hypothesis impact the power manifold in greater detail.
\end{itemize}





\section{Conclusion}

In this work, we show that PCA-derived features are beneficial for exploring the high-dimensional manifolds of the power surface. We provide multiple algorithms that provide reasonable alternatives to the original brute force method of a power analysis. We show that using a simple, fully connected neural network, we can generalize across complex and even non-linear models to consistently predict power. We showcase that even using 10\% training data can lead to high prediction accuracy for regression and classification tasks. Finally, with transfer learning, we can learn a single model and boost the performance of other models again with only 10\% training data.

\bibliography{main.bib}
\bibliographystyle{collas2022_conference}

\appendix
\section{Appendix}

\section{PNN Performance}

We demonstrate additional results about the model's performance across changes in the number of predictors, for different classification boundaries (0.6 instead of 0.8), and the performance for regression tasks.

\begin{figure}[htb!]
    \centering
    \begin{subfigure}[b]{0.45\textwidth}
    \includegraphics[width=\linewidth]{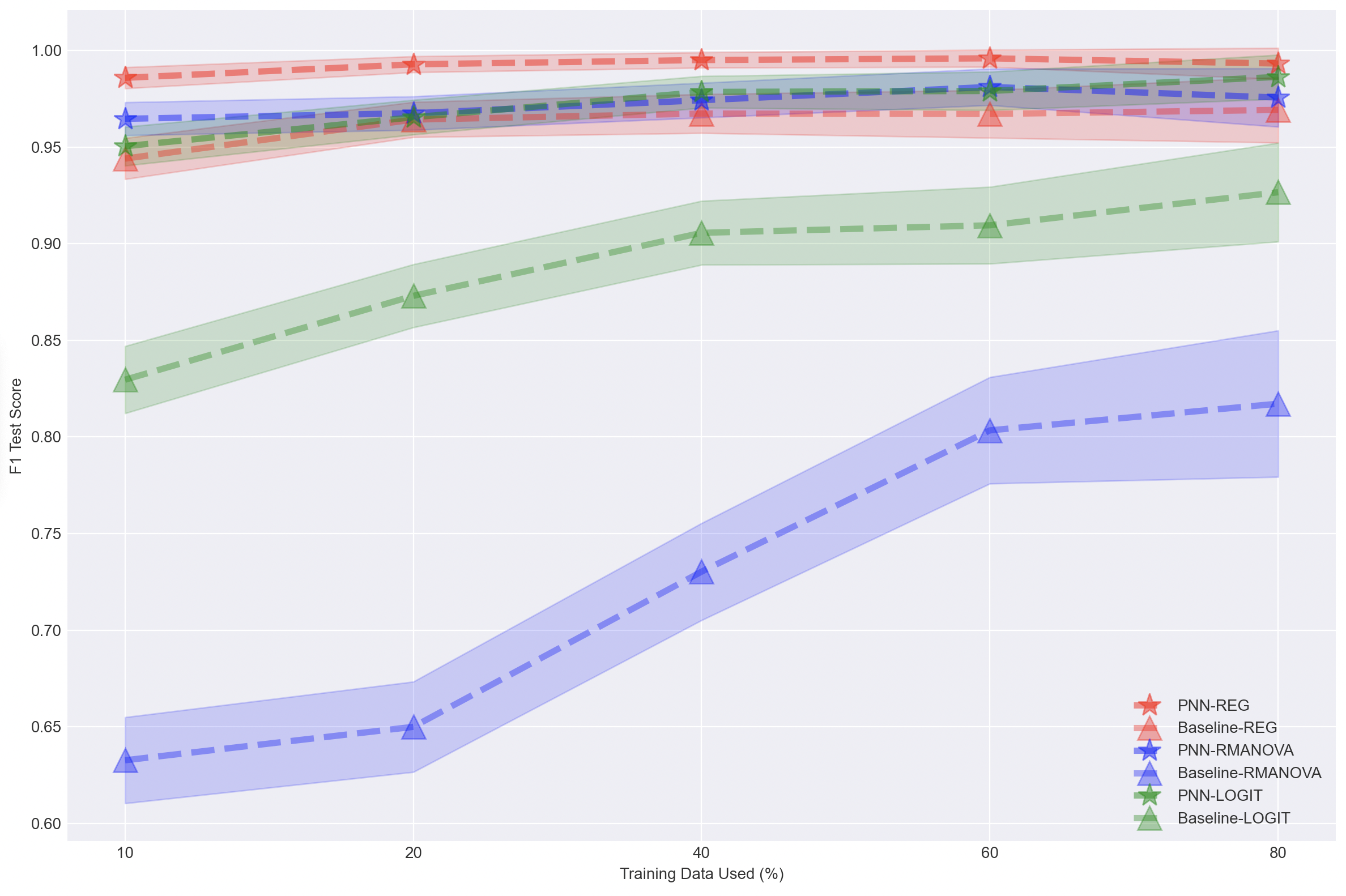}
    \caption{}\label{fig:REGR0}
    \end{subfigure}
    \begin{subfigure}[b]{0.45\textwidth}
    \includegraphics[width=\linewidth]{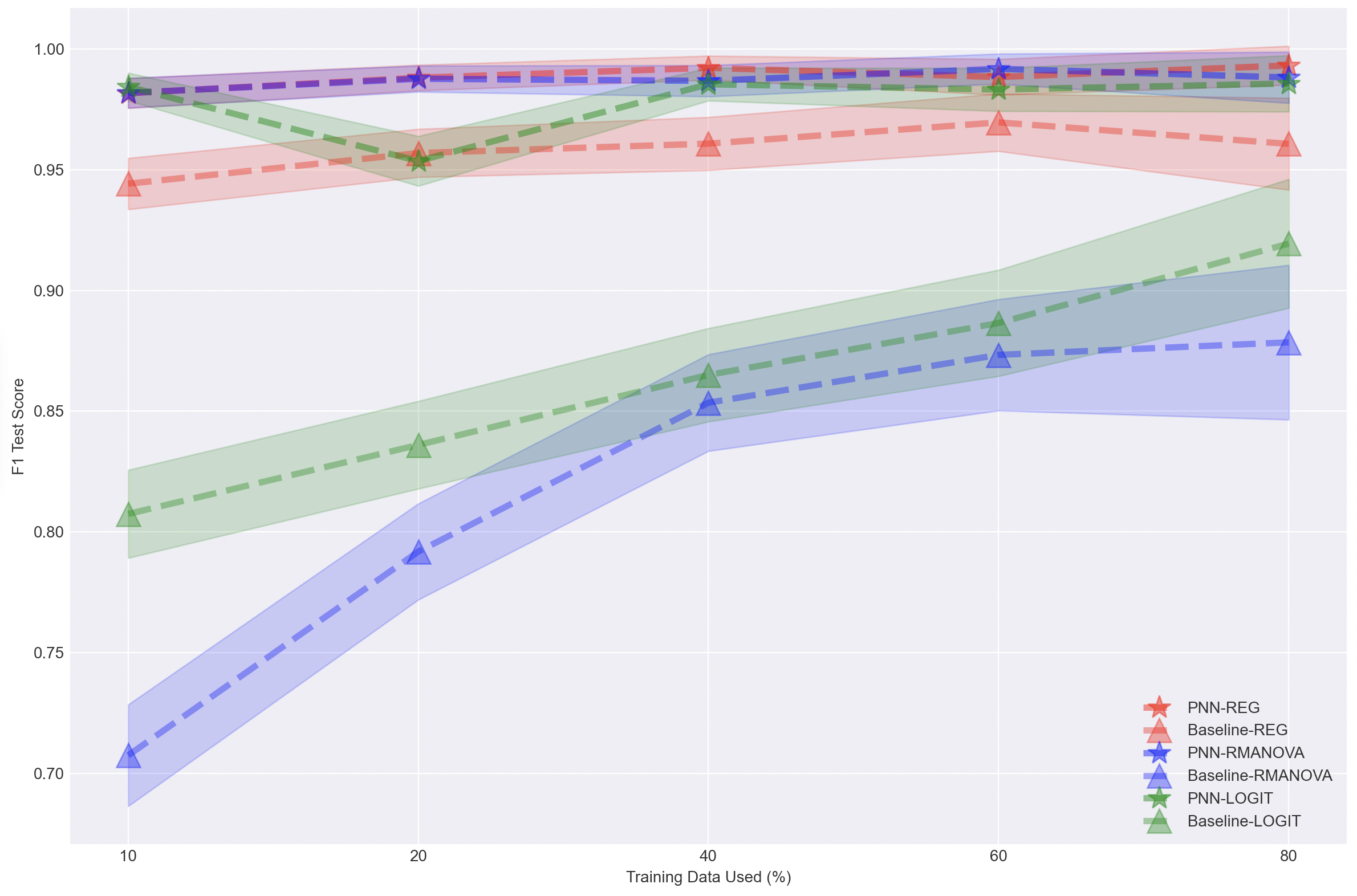}\caption{}\label{B}
    \end{subfigure}
    \caption{\textbf{PNN Classification Performance:} Comparison of model performance over change in proportion of training data used (increasing from left to right). 
    Results for the 0.6 classification boundary for \textbf{(left)}\textit{} 3 predictors , and \textbf{\textit{(right)}} 5 predictors.}
    \label{fig:PLOT61}
\end{figure}

\begin{figure}[htb!]
    \centering
    \begin{subfigure}[b]{0.45\textwidth}
    \includegraphics[width=\linewidth]{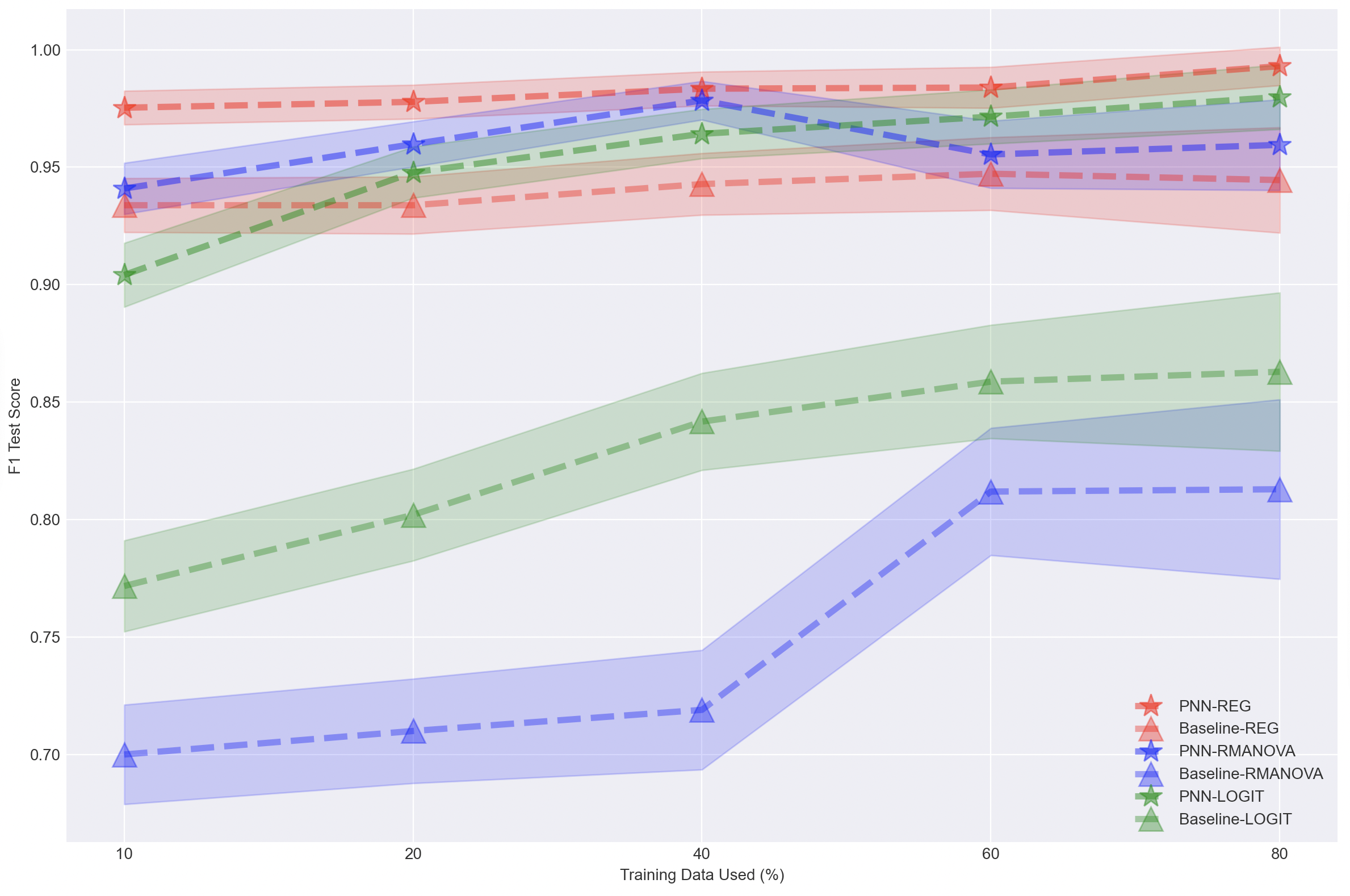}
    \caption{}\label{fig:REGR1}
    \end{subfigure}
    \begin{subfigure}[b]{0.45\textwidth}
    \includegraphics[width=\linewidth]{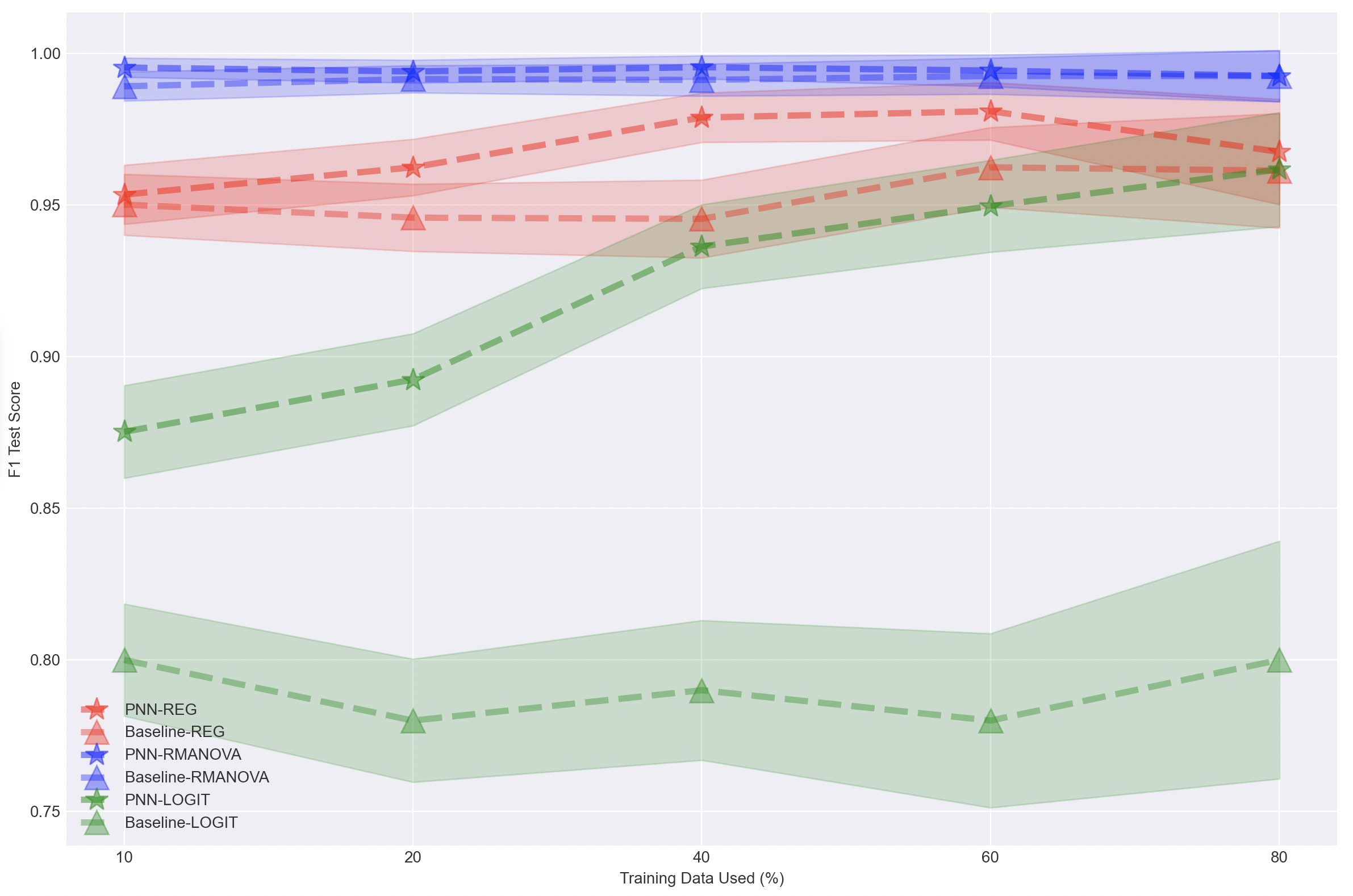}\caption{}\label{B5}
    \end{subfigure}
    \caption{\textbf{PNN Classification Performance:} Comparison of model performance over change in proportion of training data used (increasing from left to right). 
    Results for the 0.6 classification boundary for \textbf{(left)}\textit{} 10 predictors , and \textbf{\textit{(right)}} 20 predictors.}
    \label{fig:PLOT62}
\end{figure}

We first consider changes in the classification boundary from 0.8 to 0.6 to observe models with lower power as potentially interesting. We observe from Figs.~\ref{fig:PLOT61} and \ref{fig:PLOT62}.

\begin{figure}[htb!]
    \centering
    \begin{subfigure}[b]{0.45\textwidth}
    \includegraphics[width=\linewidth]{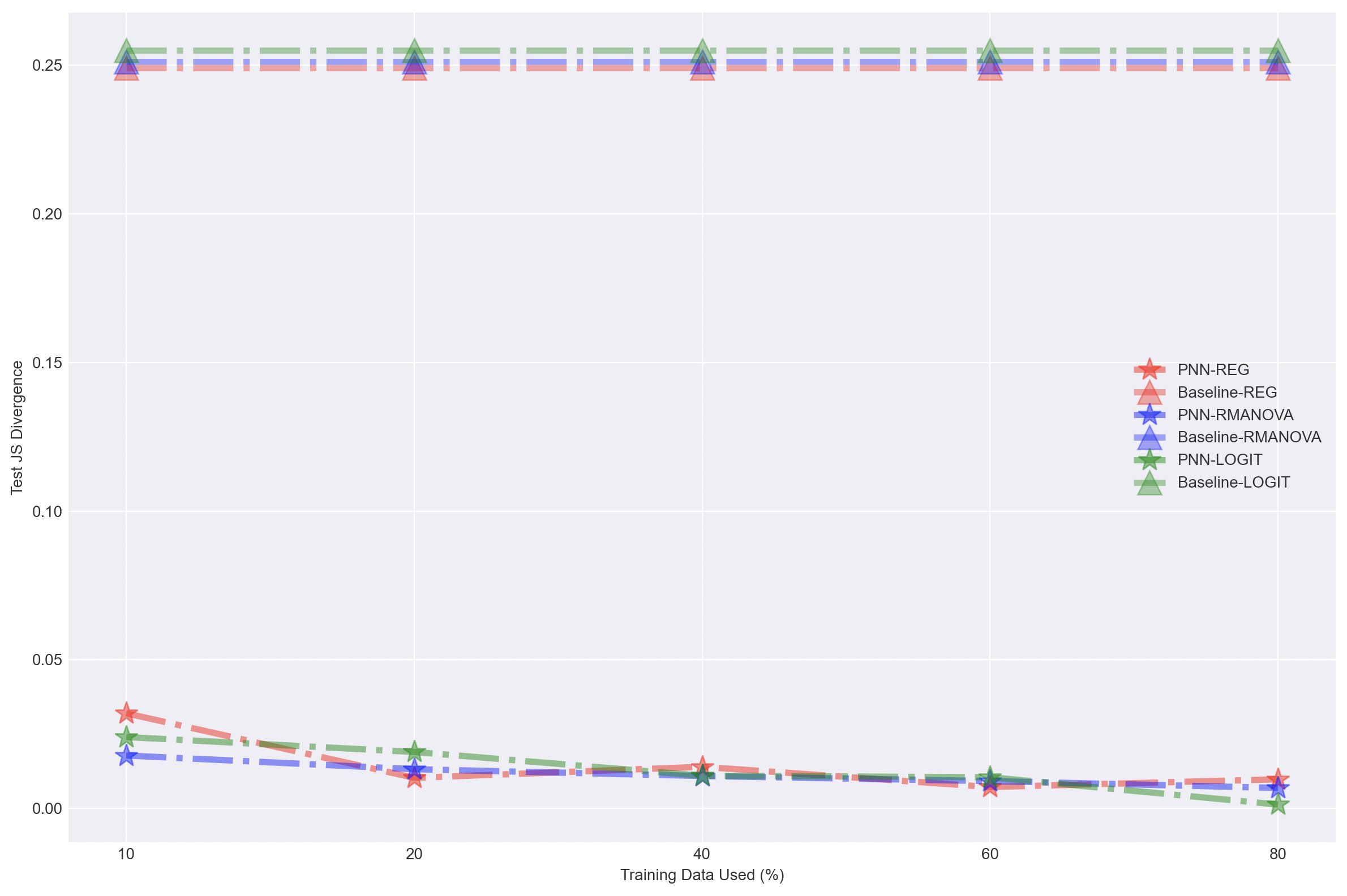}
    \caption{}\label{fig:REGR9}
    \end{subfigure}
    \begin{subfigure}[b]{0.45\textwidth}
    \includegraphics[width=\linewidth]{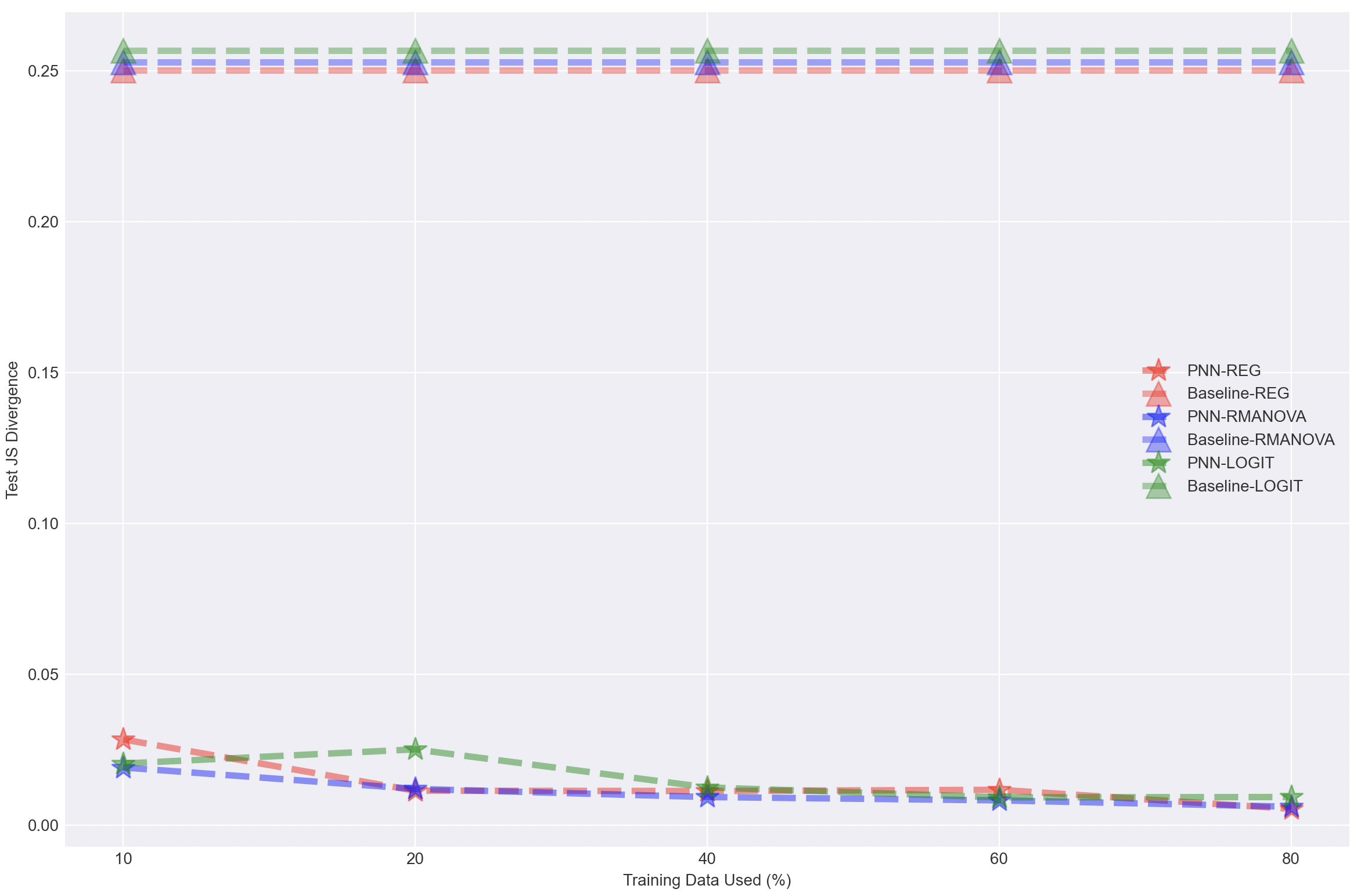}\caption{}\label{B20}
    \end{subfigure}
    \caption{\textbf{PNN REG Regression Performance:} Regression task is a super set of the classification task and thus inherently much harder. We report the performance of this task over change in the number of training points. Interestingly, even with just 10\% of the data the neural network performs close enough to the neural network using 8x training data. Results for the REG model family \textit{\textbf{(left)}} (classification boundary 0.8) and \textit{\textbf{(right)}} (classification boundary 0.6).}
    \label{fig:REGR}
\end{figure}

\begin{figure}[htb!]
    \centering
    \begin{subfigure}[b]{0.45\textwidth}
    \includegraphics[width=\linewidth]{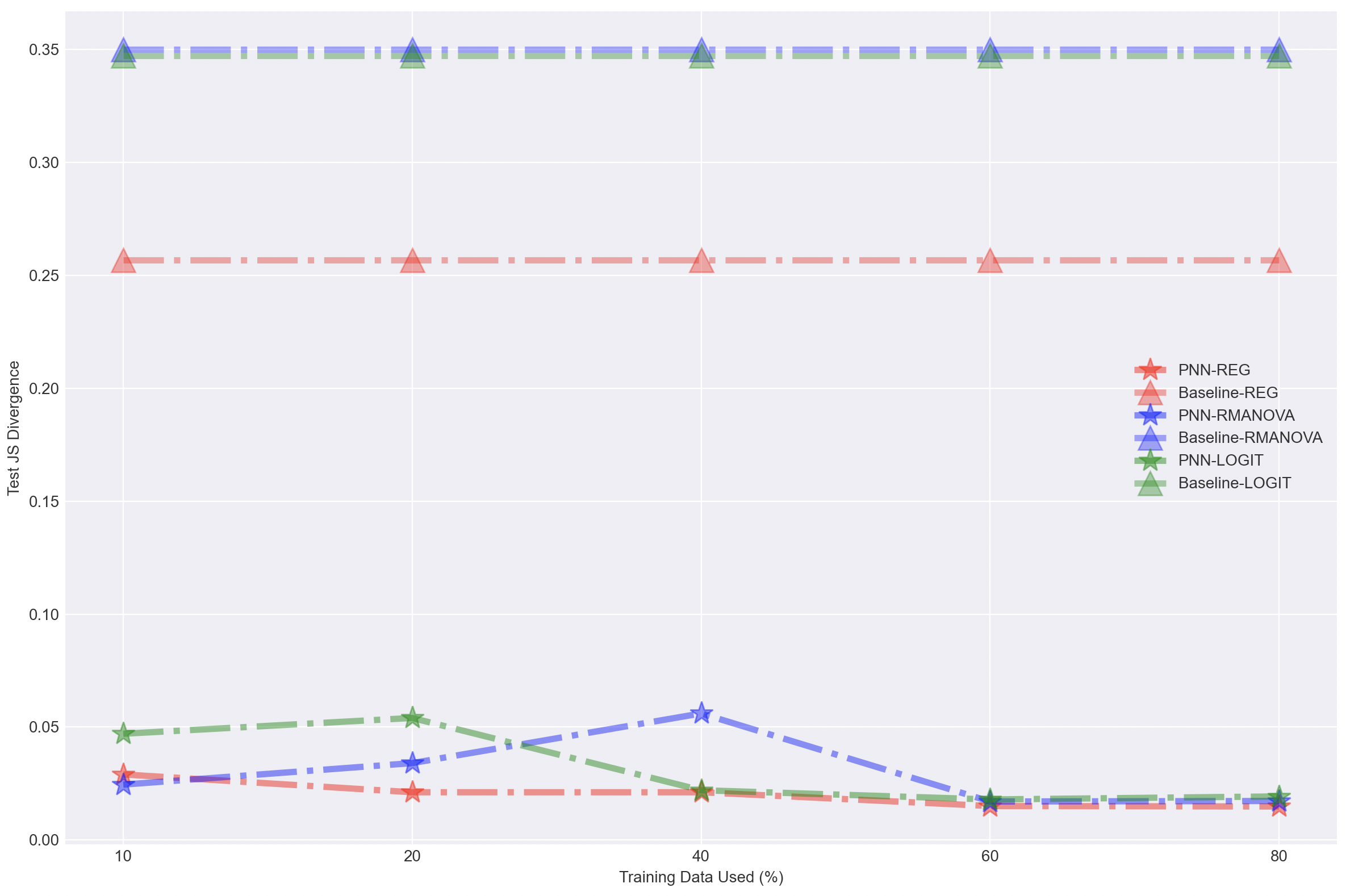}
    \caption{}\label{A1}
    \end{subfigure}
    \begin{subfigure}[b]{0.45\textwidth}
    \includegraphics[width=\linewidth]{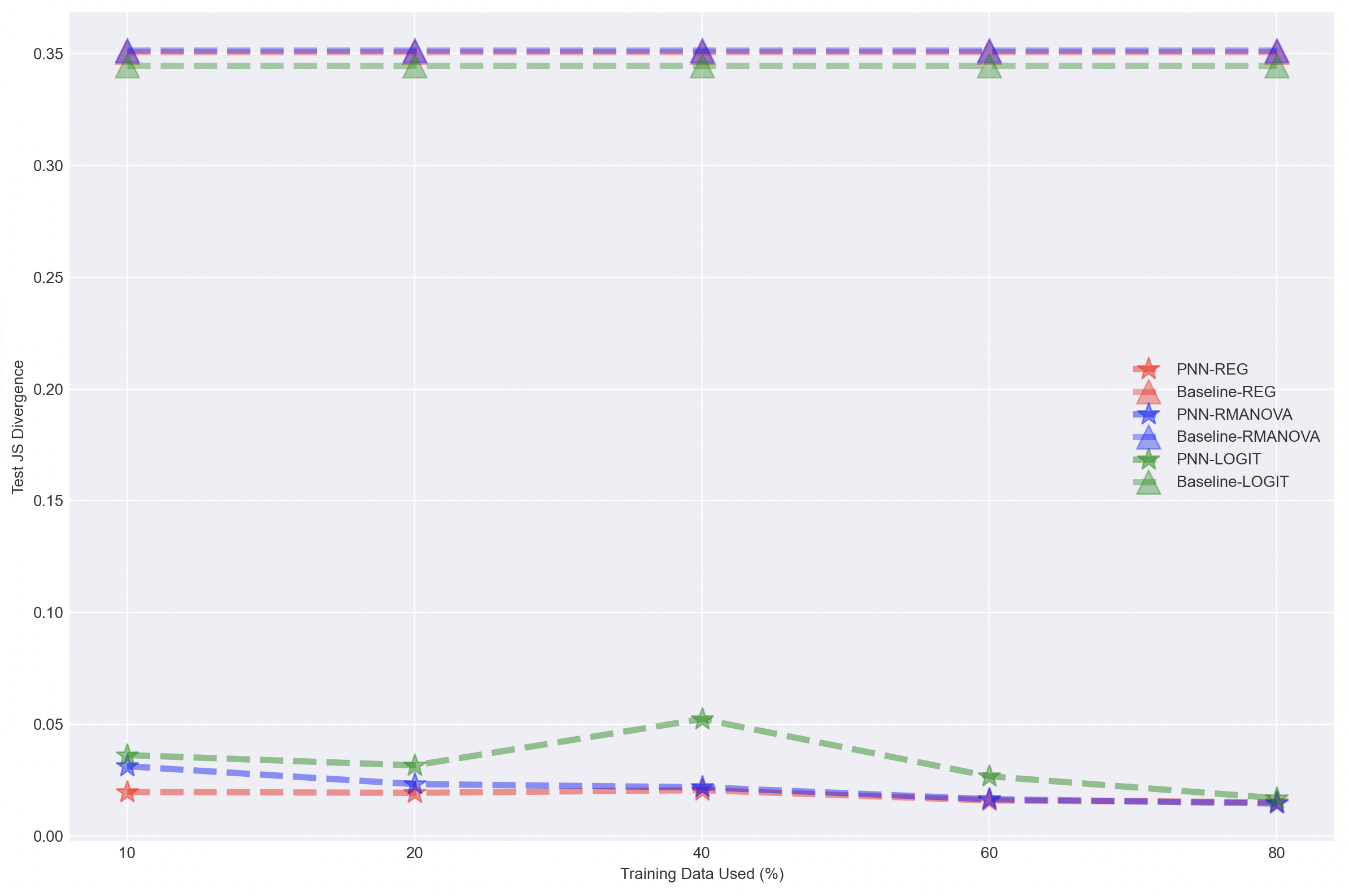}\caption{}\label{B1}
    \end{subfigure}
    \caption{\textbf{PNN RMANOVA Regression Performance:} Regression task is a super set of the classification task and thus inherently much harder. We report the performance of this task over change in the number of training points. Interestingly, even with just 10\% of the data the neural network performs close enough to the neural network using 8x training data. Results for the RMANOVA model family \textit{\textbf{(left)}} (classification boundary 0.8) and \textit{\textbf{(right)}} (classification boundary 0.6).}
    \label{fig:RMANOVAR}
\end{figure}

\begin{figure}[htb!]
    \centering
    \begin{subfigure}[b]{0.45\textwidth}
    \includegraphics[width=\linewidth]{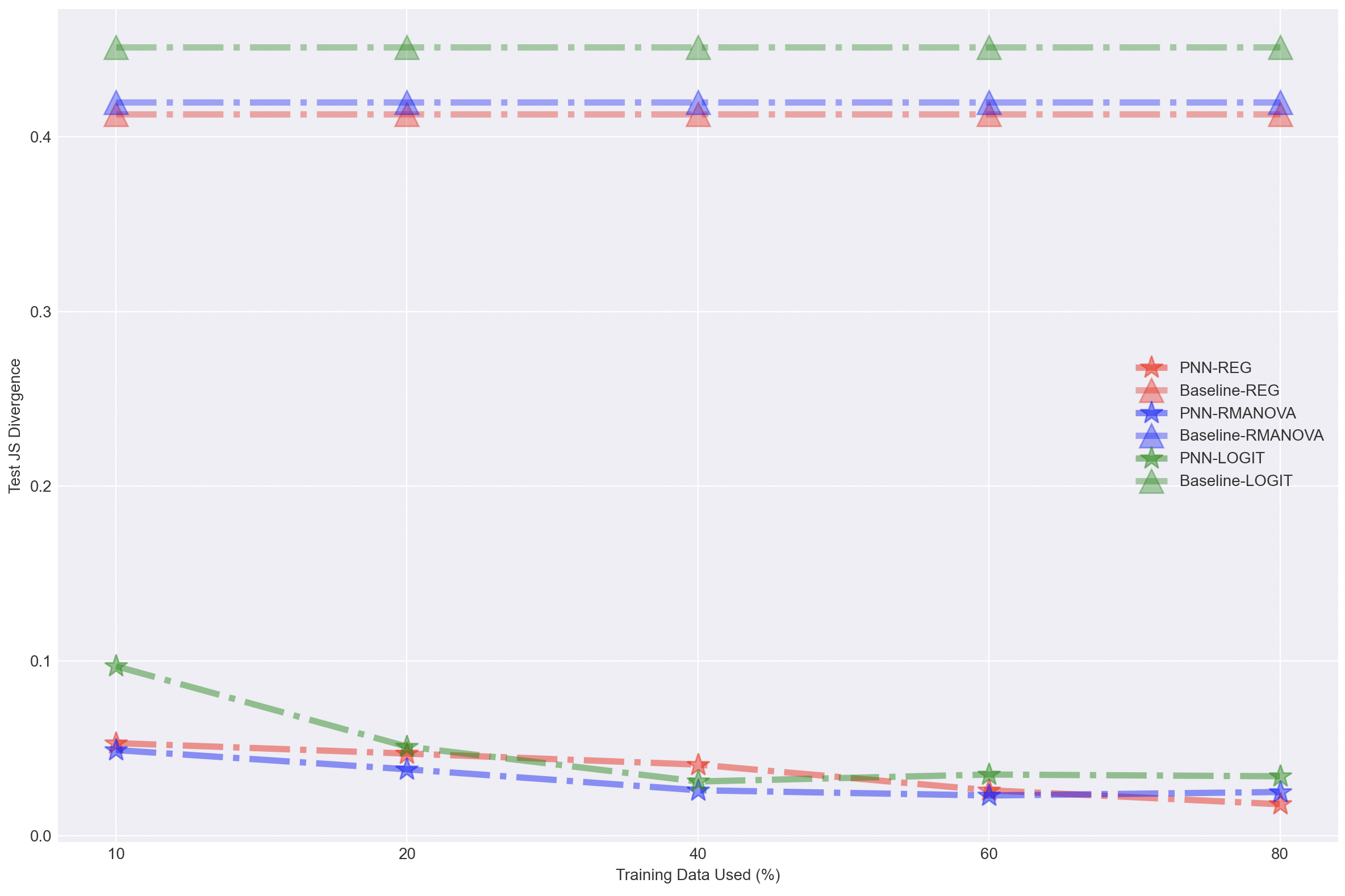}
    \caption{}\label{A5}
    \end{subfigure}
    \begin{subfigure}[b]{0.45\textwidth}
    \includegraphics[width=\linewidth]{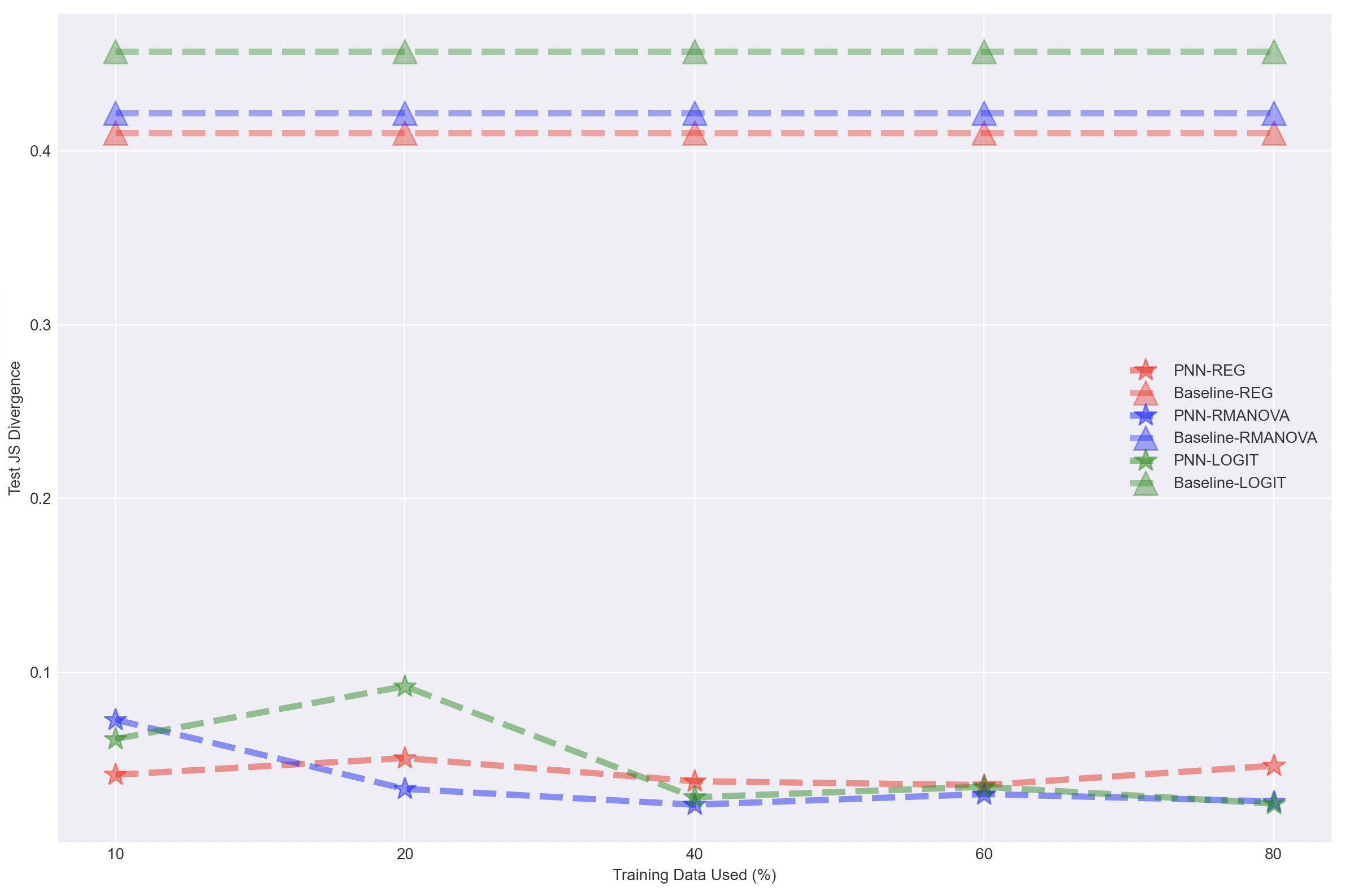}\caption{}\label{B2}
    \end{subfigure}
    \caption{\textbf{PNN LOGIT Regression Performance:} Regression task is a super set of the classification task and thus inherently much harder. We report the performance of this task over change in the number of training points. Interestingly, even with just 10\% of the data the neural network performs close enough to the neural network using 8x training data. Results for the LOGIT model family \textit{\textbf{(left)}} (classification boundary 0.8) and \textit{\textbf{(right)}} (classification boundary 0.6).}
    \label{fig:LOGITR}
\end{figure}

For all regression tasks in Figs.~\ref{fig:REGR}, \ref{fig:RMANOVAR}, \ref{fig:LOGITR} (even over the change in the classification boundary from 0.8 to 0.6) we can observe that the JS divergence over the test set is always better than the random baseline by at least 10x and going up to even 200x performance.

\begin{figure}[htb!]
    \centering
    \begin{subfigure}[b]{0.45\textwidth}
    \includegraphics[width=\linewidth]{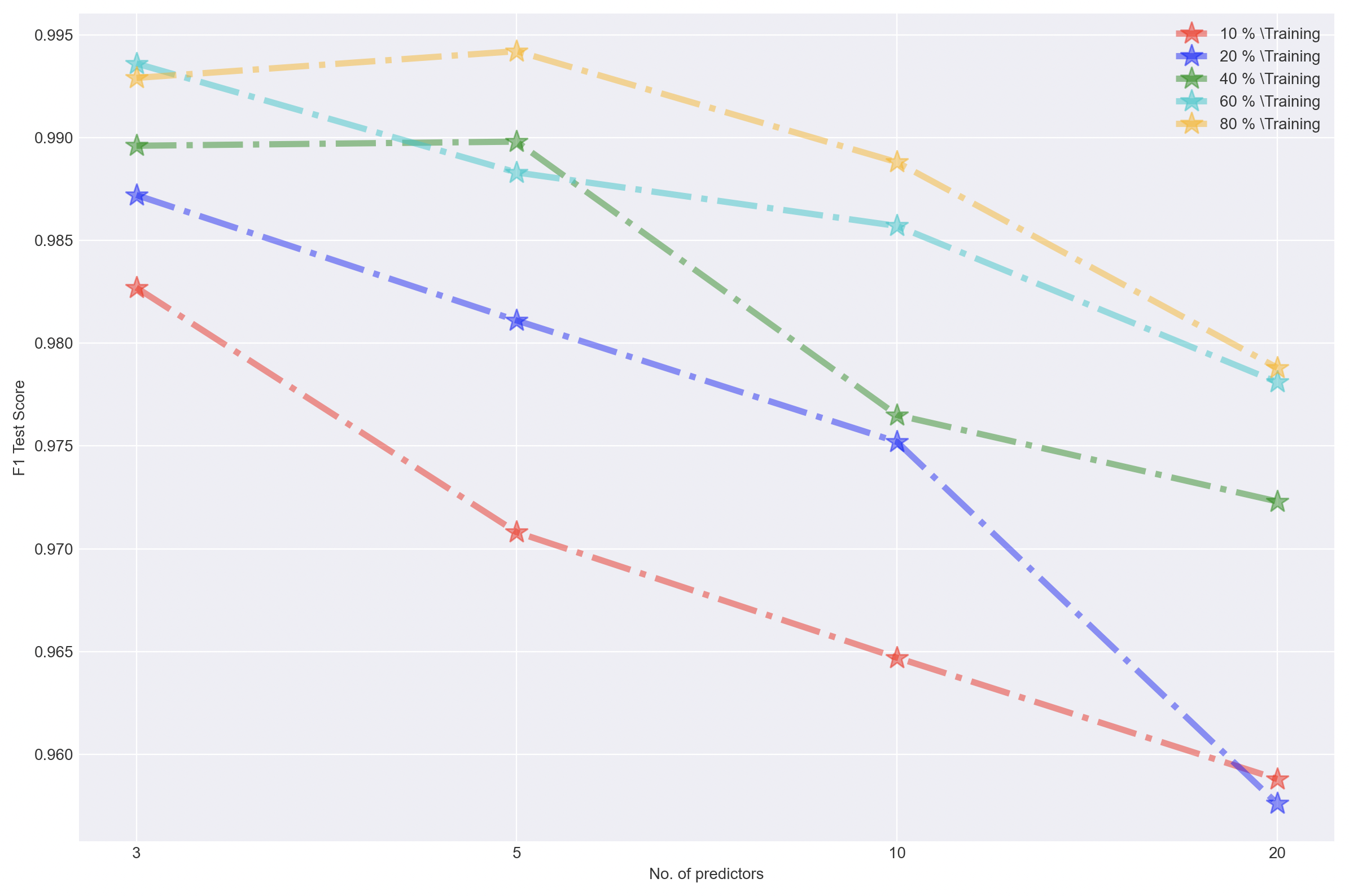}
    \caption{}\label{A2}
    \end{subfigure}
    \begin{subfigure}[b]{0.45\textwidth}
    \includegraphics[width=\linewidth]{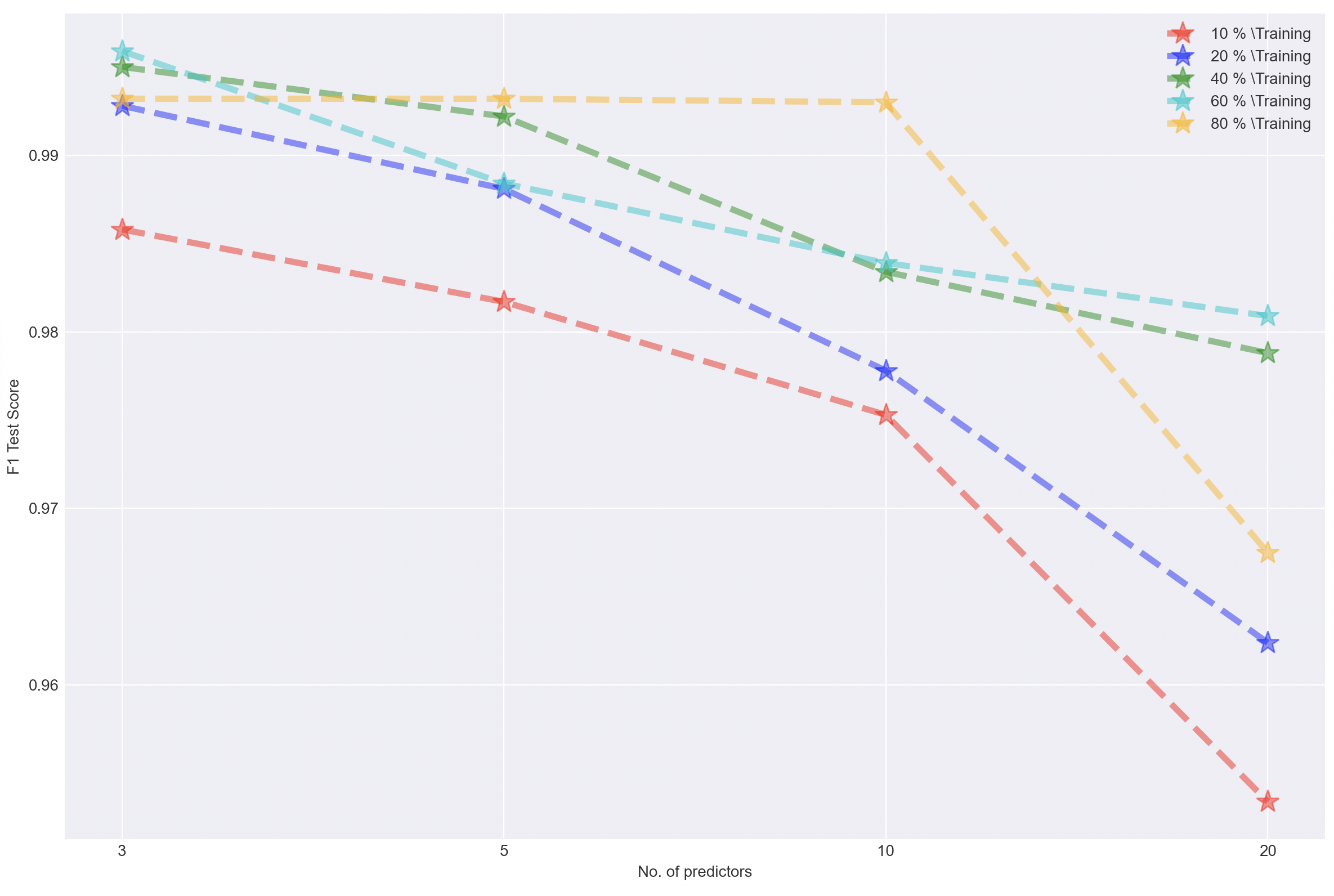}\caption{}\label{B22}
    \end{subfigure}
    \caption{\textbf{PNN REG Classification Performance:} We demonstrate how the change in model complexity due to increase in predictors affects PNN's performance. \textit{\textbf{(left)}} (classification boundary 0.8) and \textit{\textbf{(right)}} (classification boundary 0.6).}
    \label{fig:classREG}
\end{figure}

\begin{figure}[htb!]
    \centering
    \begin{subfigure}[b]{0.45\textwidth}
    \includegraphics[width=\linewidth]{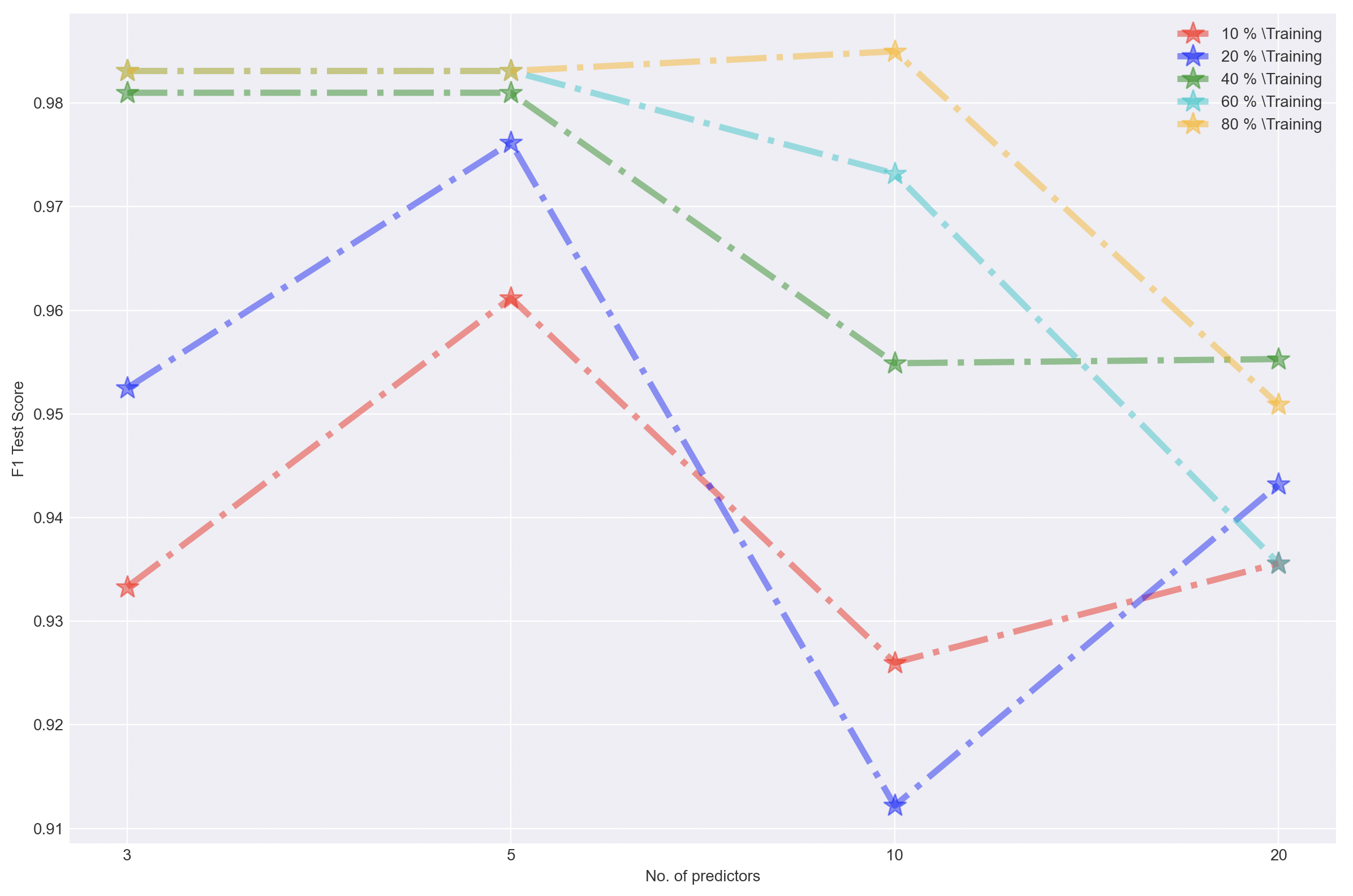}
    \caption{}\label{A3}
    \end{subfigure}
    \begin{subfigure}[b]{0.45\textwidth}
    \includegraphics[width=\linewidth]{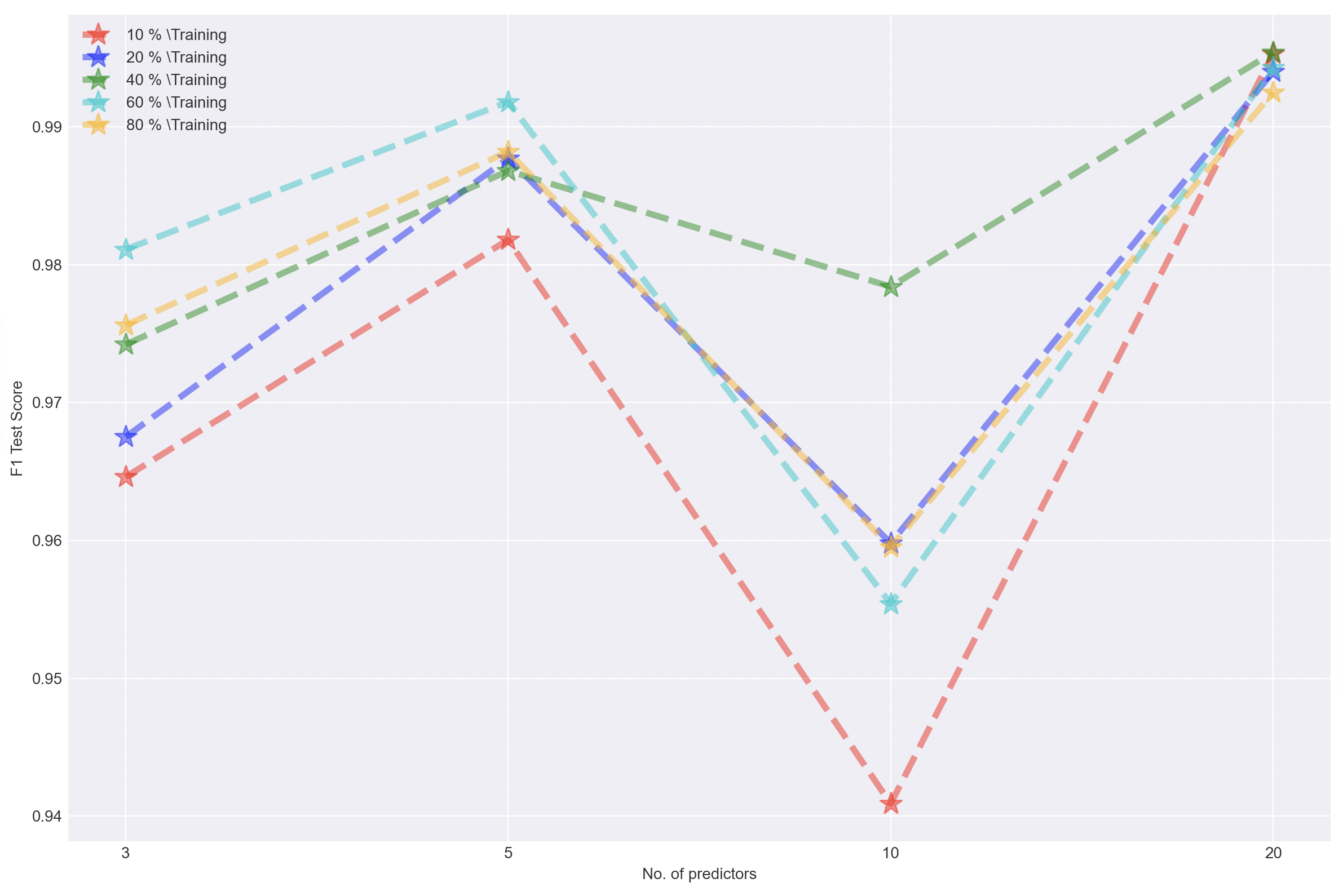}\caption{}\label{B3}
    \end{subfigure}
    \caption{\textbf{PNN RMANOVA Classification Performance:} We demonstrate how the change in model complexity due to increase in predictors affects PNN's performance. \textit{\textbf{(left)}} (classification boundary 0.8) and \textit{\textbf{(right)}} (classification boundary 0.6).}
    \label{fig:classRMANOVA}
\end{figure}

\begin{figure}[htb!]
    \centering
    \begin{subfigure}[b]{0.45\textwidth}
    \includegraphics[width=\linewidth]{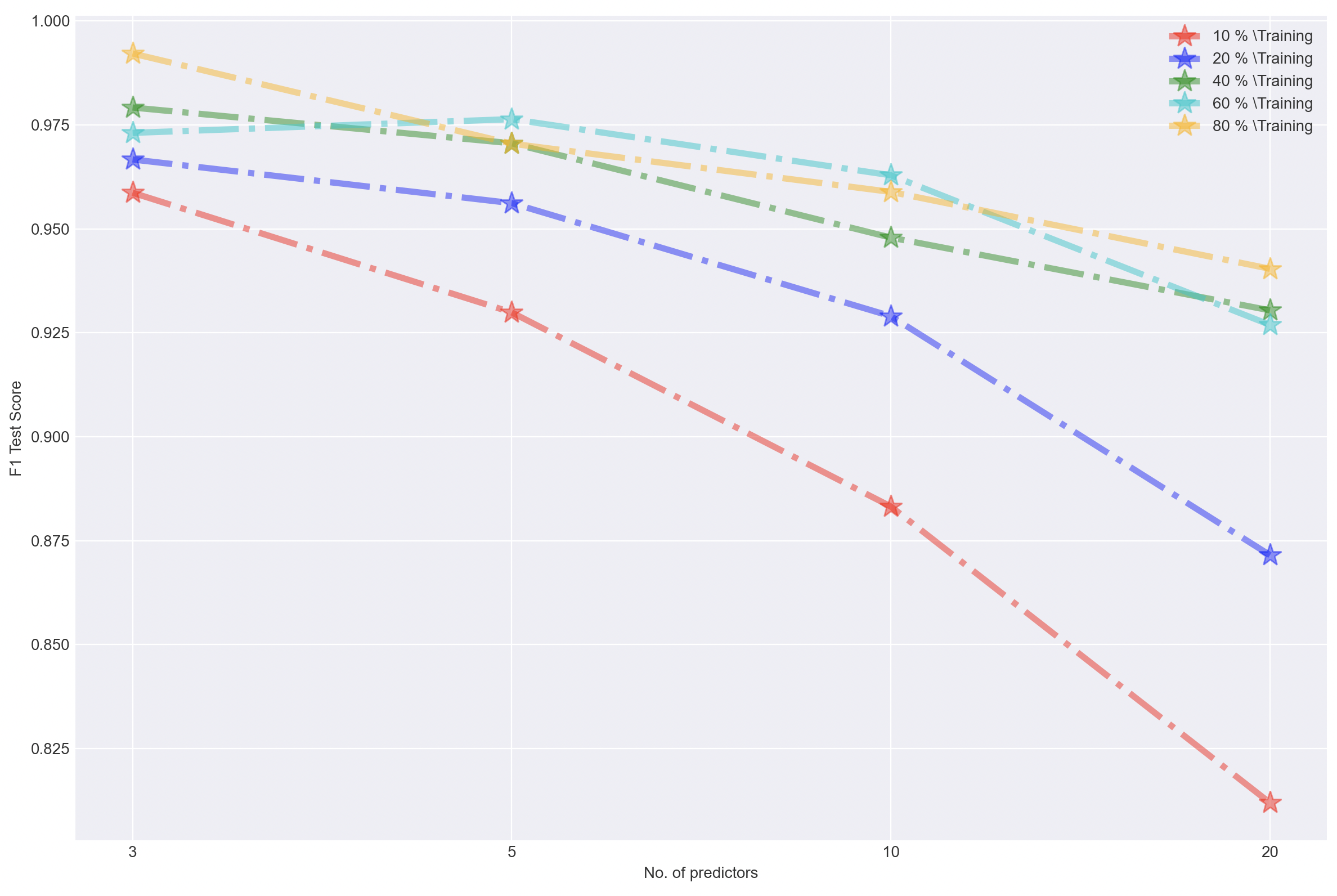}
    \caption{}\label{A6}
    \end{subfigure}
    \begin{subfigure}[b]{0.45\textwidth}
    \includegraphics[width=\linewidth]{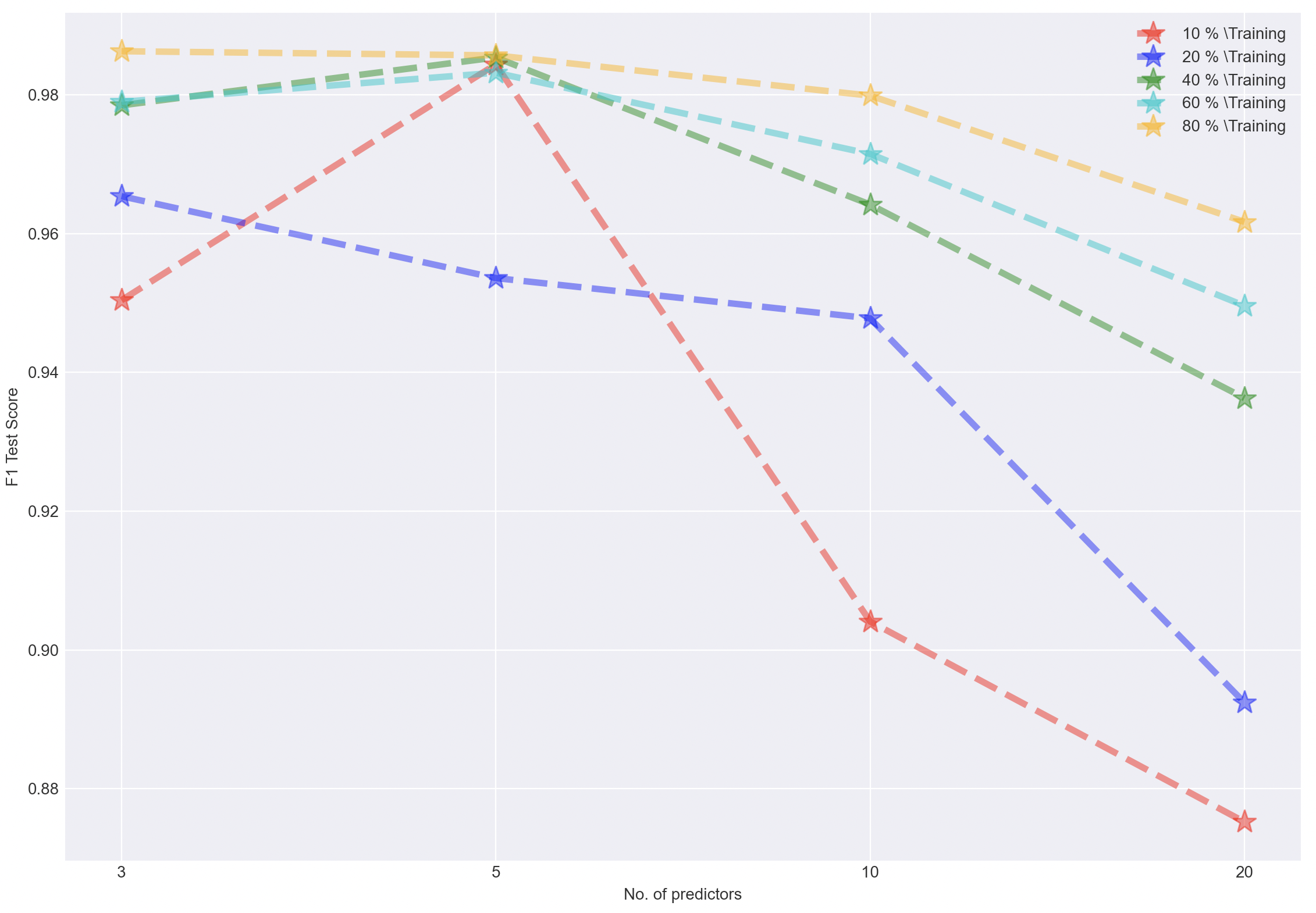}\caption{}\label{B6}
    \end{subfigure}
    \caption{\textbf{PNN LOGIT Classification Performance:} We demonstrate how the change in model complexity due to increase in predictors affects PNN's performance. \textit{\textbf{(left)}} (classification boundary 0.8) and \textit{\textbf{(right)}} (classification boundary 0.6).}
    \label{fig:classLOGIT}
\end{figure}

The performance over the change in the number of predictors as seen in Figs.~\ref{fig:classREG}, \ref{fig:classRMANOVA} and \ref{fig:classLOGIT} shows a downward trend for the REG and LOGIT models and so the classifier consistently provides poor performance with an increase in model complexity due to the number of predictors. This can be explained by the massive increase in the dimensionality of the space. However, for RMANOVA we do not see such a trend at all, and rather the models can have higher performance for even the complex models. We believe this is a current limitation due to sub-optimal model choice (not driven by the validation set).

\subsection{Parameter Tuning}\label{sec:tuning}

Tuning the algorithms used in this work is critical to improving the performance of the Power Network and Power Cluster. It turns out that choosing the variance captured by PCA directly affects the Power Network. Moreover, PCA variance and the number of clusters selected in Power Clustering can improve its performance.

\subsubsection{Experiment Details}\label{section:experiments}

We have standardized the parameter distribution for data collection to provide a consistent comparison. These parameters can be found in Table~\ref{table:exp-guidance}. For tuning the neural network's learning rate, we use the library Optuna (\cite{akiba2019optuna}). 

\begin{table}[htb!]
\begin{center}
\begin{tabular}{|c|c|} 
 \hline
 \rowcolor{lightgray}
 Experimentation Details & Values \\ 
 \hline \hline
 No. of power simulations for each record & 1000 \\
 \rowcolor{lightgray}
 Total Datapoints per model & 2000 \\ 
 No. of training epochs & 500 \\ 
 \rowcolor{lightgray}
 Training Data Split & 10, 20, 40, 60, 80 (\%)\\
 Power Classification Boundary & 0.6, 0.8 \\
 \rowcolor{lightgray}
 Optimizer & Adam \\
 Model Performance confidence interval & 95\% \\
 \hline
\end{tabular}
\end{center}
\caption{Experimental Parameters}
\label{table:exp-guidance}
\end{table}

\subsubsection{PCA}

\begin{figure}[htb!]
     \centering
    \includegraphics[width=0.5\columnwidth]{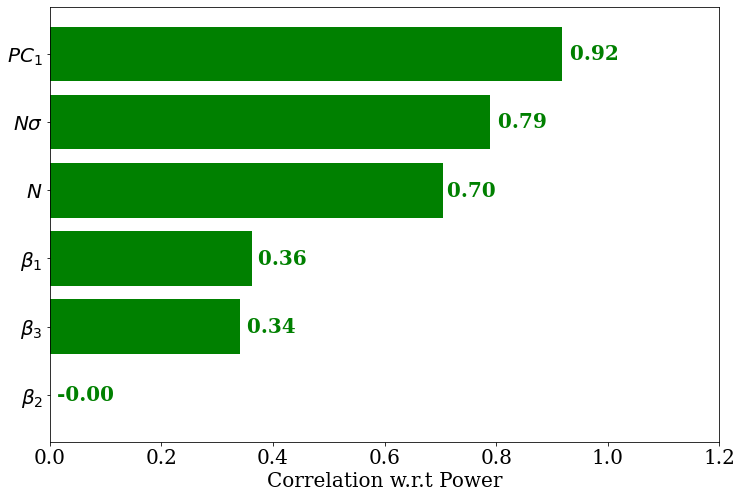}
    \caption{\textbf{Impact of PCA: }Correlation of parameters with \textit{true} power. With feature engineering and PCA we are able to obtain correlation $> 0.9$ for a linear regression model with 3 parameters.}
    \label{fig:corr}
\end{figure}

Usually, increasing the sample size improves power. Furthermore, our feature engineered dimension denoted by $N \sigma$ \textit{\emph{i.e., }(scaled weight)} also affects the power significantly. We can observe the correlation values concerning power in Fig.~\ref{fig:corr}. Thus, we expect the variance covered in PCA transformed data to require at least two dimensions.

We consistently observe excellent performance after selecting 99\% variance from PCA. Furthermore, we concatenate these new features to the original matrix and use this to train the Neural Network. As we see from figure~\ref{fig:corr}, the impact of the PCA features is sizeable.

\subsubsection{Choice of clusters in Power Cluster}

Depending on how the data is clustered, the Choice of several clusters can directly impact the classification performance. However, empirically this method is unreliable and requires a visual inspection of the dataset to be helpful.

\begin{figure}[htb!]
     \centering
    \includegraphics[width=\columnwidth]{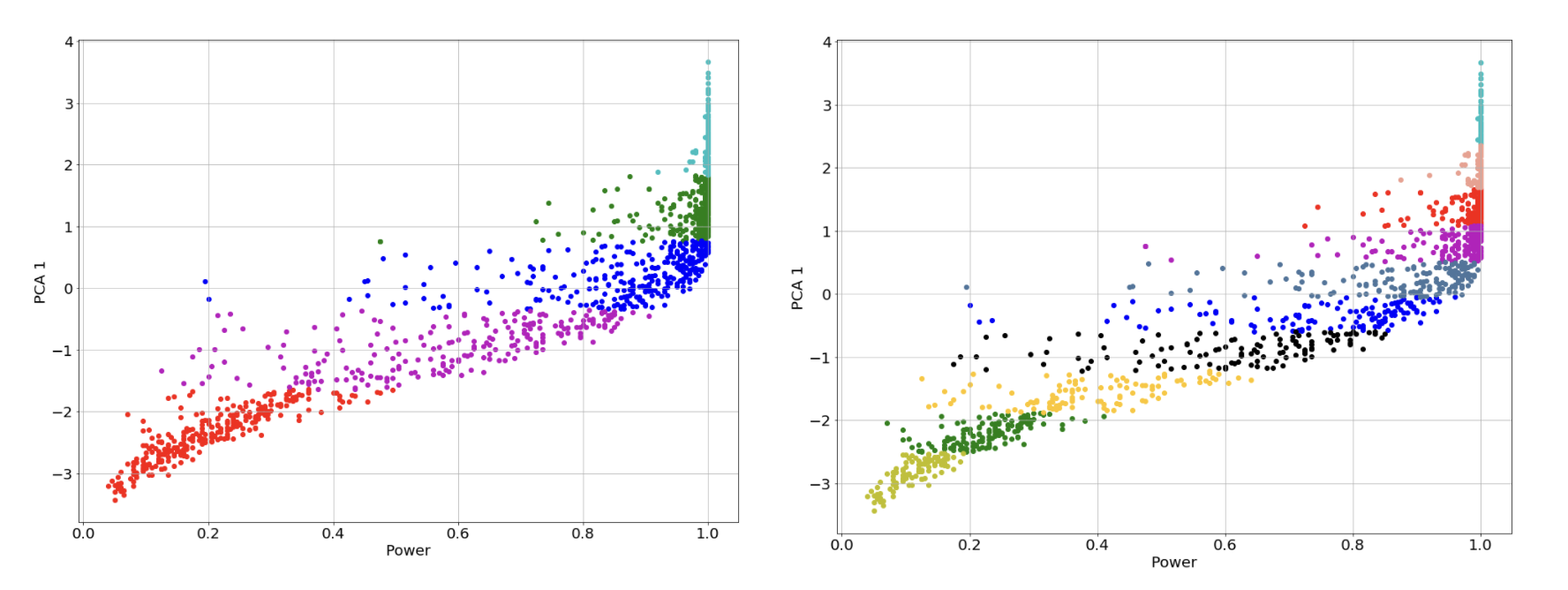}
    \caption{\textbf{P-CLUSTER Performance (multi-clustering):} P-CLUSTER can identify multiple zones in the power domain quite efficiently. \textit{(left)} displays 5 clusters and, \textit{(right)} displays 10 clusters.}
    \label{fig:mcluster}
\end{figure}

\subsubsection{Other Considerations}

We employ a few techniques during neural network training with increasingly complex models and parameter spaces.
\begin{itemize}
    \item The training is sensitive to the Choice of the learning rate. Tuning needs to be achieved after using a hyperparameter tuner like Optuna \cite{akiba2019optuna}.
    \item Training tends to fluctuate a lot, and a higher number of epochs combined with a validation set and early stopping seem to help.
\end{itemize}

\subsubsection{Neural Network Architecture}

\begin{table}[htb!]
\begin{center}
\begin{tabular}{|c|c|} 
 \hline
 \rowcolor{lightgray}
 Layer & Parameters \\ 
 \hline \hline
 Dense (input) & 64 units + ReLU \\
 \rowcolor{lightgray}
 Dense (hidden) & 32 units + ReLU\\
 Dense (output) & 1 unit + Sigmoid\\
 \hline
\end{tabular}
\end{center}
\caption{\textit{PNN} Architecture: We report the fully connected layers. The input dimensions depend on the number of predictors and additional PCA features.}
\label{table:pnn-arch}
\end{table}

\subsubsection{List of Model Parameters}
\begin{table}[htb!]
\begin{center}
\scalebox{0.8}{
\begin{tabular}{|c|c|c|c|c|} 
 \hline
 \rowcolor{lightgray}
 $D_{X_{1}}$ & $D_{X_{2}}$ & $D_{X_{3}}$ & $D_{X_{4}}$ & $D_{X_{5}}$ \\ 
 CAT[-1, 1] & $\mathcal{N}(0, 1)$ & $X_{1} \times X_{2}$ & $\mathcal{N}(0, 2)$ & $X_{4} \times X_{2}$\\ 
 \hline \hline
 \rowcolor{lightgray}
 $D_{X_{6}}$ & $D_{X_{7}}$ & $D_{X_{8}}$ & $D_{X_{9}}$ & $D_{X_{10}}$ \\ 
 CAT[0, 1, 2] & $\mathcal{N}(0, 2)$ & $X_{6} \times X_{7}$ & $\mathcal{N}(0, 1)$ & $X_{2} \times X_{6}$
 \\ 
 \hline \hline
 \rowcolor{lightgray}
 $D_{X_{11}}$ & $D_{X_{12}}$ & $D_{X_{13}}$ & $D_{X_{14}}$ & $D_{X_{15}}$ \\ 
$\mathcal{N}(0, 3)$ & $\mathcal{N}(0, 1)$ & $X_{1} \times X_{11}$ & $\mathcal{N}(0, 2)$ & $X_{11} \times X_{12}$
 \\ 
  \hline \hline
 \rowcolor{lightgray}
 $D_{X_{16}}$ & $D_{X_{17}}$ & $D_{X_{18}}$ & $D_{X_{19}}$ & $D_{X_{20}}$ \\ 
$\mathcal{N}(0, 2)$ & $\mathcal{N}(0, 2)$ & $X_{11} \times X_{14}$ & $\mathcal{N}(0, 1)$ & $X_{6} \times X_{16}$ \\ \hline 
\end{tabular}}
\end{center}
\caption{$D_{\mathcal{O}}$: Distribution of model features. For predictors $k < 20$, we only select the first $k$ predictors. Note that CAT refers to a categorical variable while $\mathcal{N(\mu, \sigma)}$ is the Normal Distribution with mean $\mu$ and standard deviation $\sigma$.}
\label{table:distribution-original}
\end{table}

\begin{table}[htb!]
\begin{center}
\scalebox{0.8}{
\begin{tabular}{|c|c|c|c|c|} 
 \hline
 \rowcolor{lightgray}
 $D_{X_{1}}$ & $D_{X_{2}}$ & $D_{X_{3}}$ & $D_{X_{4}}$ & $D_{X_{5}}$ \\ 
 $\mathcal{N}(0, 1)$ & CAT[-1, 1] & $\mathcal{N}(0, 1)$ & $\mathcal{N}(0, 1)$ & $X_{2} \times X_{1}$\\ 
 \hline \hline
 \rowcolor{lightgray}
 $D_{X_{6}}$ & $D_{X_{7}}$ & $D_{X_{8}}$ & $D_{X_{9}}$ & $D_{X_{10}}$ \\ 
 $X_{2} \times X_{3}$ & $X_{2} \times X_{4}$ & $\mathcal{N}(0, 1)$ & $\mathcal{N}(0, 1)$ & $X_{2} \times X_{8}$
 \\ 
 \hline \hline
 \rowcolor{lightgray}
 $D_{X_{11}}$ & $D_{X_{12}}$ & $D_{X_{13}}$ & $D_{X_{14}}$ & $D_{X_{15}}$ \\ 
$\mathcal{N}(0, 3)$ & $\mathcal{N}(0, 1)$ & $X_{1} \times X_{11}$ & $\mathcal{N}(0, 2)$ & $X_{11} \times X_{12}$
 \\ 
  \hline \hline
 \rowcolor{lightgray}
 $D_{X_{16}}$ & $D_{X_{17}}$ & $D_{X_{18}}$ & $D_{X_{19}}$ & $D_{X_{20}}$ \\ 
$\mathcal{N}(0, 2)$ & $\mathcal{N}(0, 2)$ & $X_{11} \times X_{14}$ & $\mathcal{N}(0, 1)$ & $X_{6} \times X_{16}$ \\ \hline 
\end{tabular}}
\end{center}
\caption{$D_{\mathcal{A}}$: Alternate Distribution of model features (used for non-linear models and testing purposes). For predictors $k < 20$, we only select the first $k$ predictors. Note that CAT refers to a categorical variable while $\mathcal{N(\mu, \sigma)}$ is the Normal Distribution with mean $\mu$ and standard deviation $\sigma$.}
\label{table:distribution-alternative}
\end{table}

\section{Computational Costs}

\begin{figure}[htb!]
    \centering
    \includegraphics[width=\columnwidth]{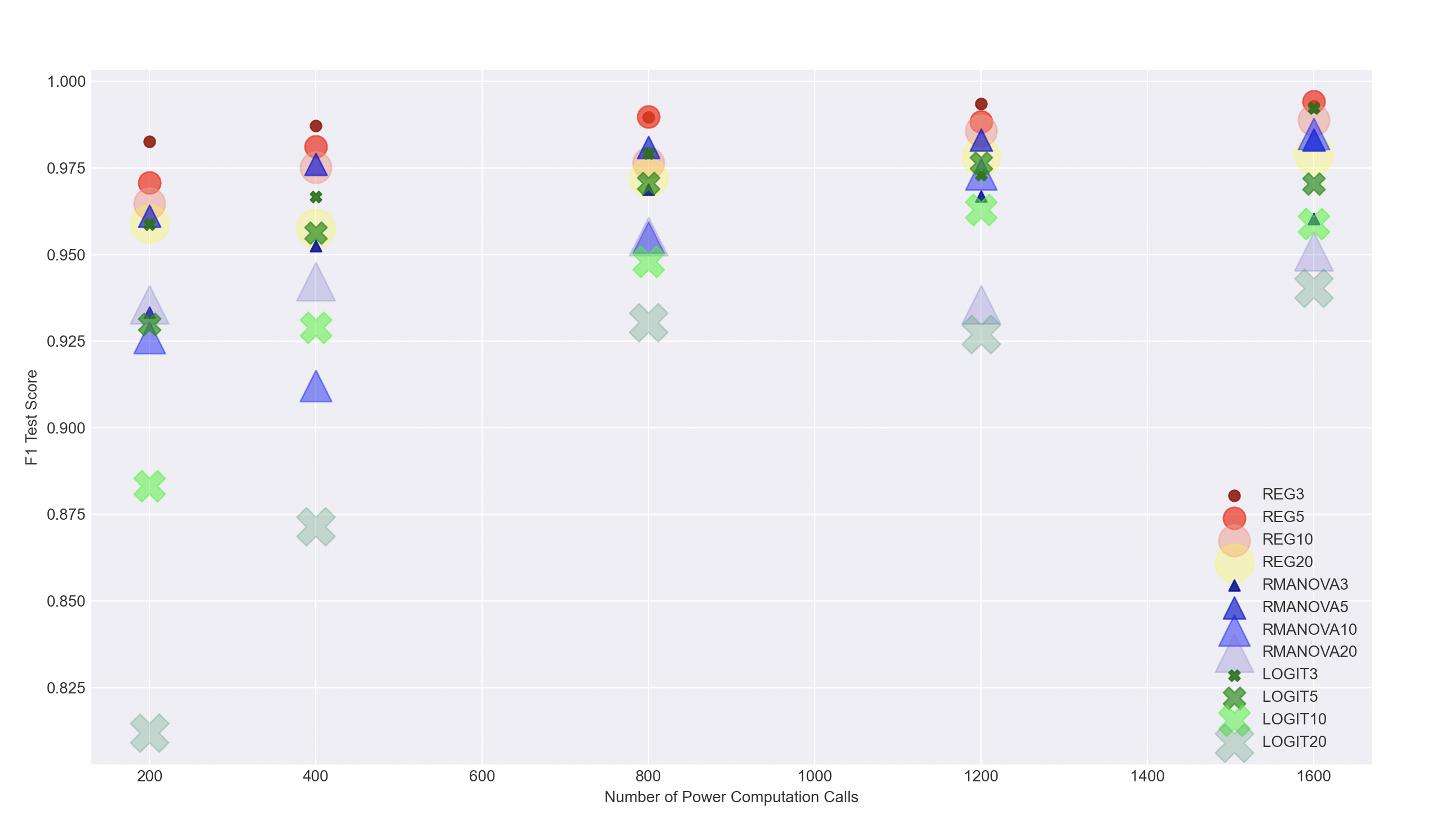}
    \caption{Cost of Power Computation: We report the calls required to compute the power manifold over a subset of the parameter space. Visualizing from left to right, we can see that we can yield high-performing classifiers even with a fraction of the training data.}
    \label{fig:comp-costs}
\end{figure}

\subsection{Details about Principal Component Analysis}

PCA works by finding the direction with maximal variance. Next, it finds the direction with the maximal variance such that this direction is uncorrelated with the previous direction. We continue in this fashion so that any pair of directions are uncorrelated. We then reorient our dataset with these new components to compute our new dataset. Suppose our original dataset is $X \in \mathbb{R}^{N \times p}$ and $v \in \mathbb{R}^{p \times k}$ are the derived Principal Components (PCs) where $N$ is the total samples in $X$, $p = $dim($X$) and $k$ is the number of PCs to be selected. Then the new dataset is given by,
\begin{equation*}
    X^{'} = X \times v.
\end{equation*}

\begin{algorithm}[htb!]
\caption{Transform the dataset into the first $k$ Principal Components}
\label{algo:pca}
\begin{algorithmic}[1]
\State \textbf{Input:} $X \in \mathbb{R}^{N \times p}$ dataset, number of samples in $X$-$N$, data dimension-$p$, number of components to be included with PCA-$k$ ($k < p$)
\State \textbf{Output:} Transformed PCA dataset
\Procedure{PCA-FIT-TRANSFORM}{$X, k$}
\For{$X_i$ in columns($X$)}
    \State $\mu_{X_{i}}$ $\leftarrow$ MEAN($X_{i}$)
    \State $\sigma_{X_{i}}$ $\leftarrow$ STANDARD-DEVIATION($X_{i}$)
    \State $X_{i}$ $\leftarrow$ $\frac{(X_{i} - \mu_{X_{i}})}{\sigma_{X_{i}}}$
\EndFor 
\State $Y \leftarrow X.T \times X$
\State $V_{k} \leftarrow$ FIRST-k-EIGENVECTORS($Y$)
\State return $X \times V_{k}$
\EndProcedure
\end{algorithmic}
\end{algorithm}

\end{document}